# Quasi-Superactivation of Classical Capacity of Zero-Capacity Quantum Channels

Laszlo Gyongyosi[*], Sandor Imre

[1]Department of Telecommunications, Budapest University of Technology and Economics

H-1111 Budapest, Magyar tudosok krt. 2. Hungary

[*]*gyongyosi@hit.bme.hu*

One of the most surprising recent results in quantum Shannon theory is the superactivation of the quantum capacity of a quantum channel. This phenomenon has its roots in the extreme violation of additivity of the channel capacity and enables to reliably transmit quantum information over zero-capacity quantum channels. In this work we demonstrate a similar effect for the classical capacity of a quantum channel which previously was thought to be impossible. We show that a nonzero classical capacity can be achieved for all zero-capacity quantum channels and it only requires the assistance of an elementary photon-atom interaction process - the stimulated emission.

# 1 Introduction

There are many phenomena in quantum systems that have no classical equivalent (or counterpart), such as entanglement, which makes it possible to store quantum information in the quantum correlation of quantum states. Quantum entanglement was discovered in the 1930s, and it may still yield many surprises in the future. It plays a fundamental role in advanced quantum communications, such as teleportation, quantum cryptography, and quantum communication processes [15-16] [20]. The characterization of quantum entanglement has deep relevance in Quantum Information Theory. Quantum entanglement is the major phenomenon which distinguishes the classical world from the quantum one. By means of entanglement, many classically totally unimaginable results can be achieved in Quantum Information Theory.

In the first decade of the 21st century, many revolutionary properties of quantum channels were discovered [1-10]. At the dawn of this millennium new problems have arisen with many open questions, which have opened the door to many new promising results such as *superactivation* of quantum channels. The *superactivation* of zero-capacity quantum channels [6-8] [13] [15] makes possible to use zero-capacity quantum channels for communication. A quantum channel can be used to realize classical information transmission or to transmit quantum information, such as quantum entanglement [11-17]. Information transmission also can be approached using the question of whether entanglement plays a role in the encoding/decoding process. This leads us to say that for quantum channels, many new capacity definitions exist in comparison to a classical communication channel [18-21], [22-25]. In the case of a classical channel, we can send only classical information.

Initially, the superactivation property was proven for the transmission of *quantum* information [13] over zero-capacity quantum channels. In this combination, each quantum channel has zero quantum capacity individually; however, their joint quantum capacity is strictly greater than zero. On the other hand, these results did not allow transmitting classical information over combination of quantum channels in which each channel has zero classical capacities. In this paper we give the mathematical proof that by the adding of quantum entanglement, two quantum channels each with zero classical capacity can be combined together to transmit classical information.

The theoretical background of the superactivation of quantum channel capacities is currently an open question; however, it is known that it is based on the extreme violation of the additivity property—in other words it is based on the non-additivity of the various quantum channel capacities with *entangled* input states. Entanglement among the input states is a required condition for the superactivation of quantum capacity of quantum channels. As we demonstrate here, a similar effect can be obtained for the classical capacity of quantum channels and entangled inputs will also have crucial importance in the proposed quasi-superactivation effect.

Sending *classical information* over a channel combination in which each channel has zero classical capacity seemed to be *impossible*. The transmission of classical information over zero-capacity quantum channels seemed to be the hardest problem among all, and it also had roots in Hastings' counterexample. As was found in 2009 [9], the Holevo information is non-additive in general; on the other hand, this result did not give an answer for the general case of the classical capacity, and it left open so many new questions. It was found that in some very special cases *quantum information* can be transmitted in a similar scenario [13-14], however the most general question — the transmission of *classical information*—over such a structure was an unsolvable problem.

The phenomenon we propose in this work is called *quasi-superactivation*. The result is similar to the superactivation effect— positive capacity can be achieved with noisy quantum channels that were initially completely useless for communication. An important difference that quasi-superactivation is limited neither by any preliminary conditions on the initial private capacity of the channel nor on the maps of other channels involved to the joint channel structure. Quasi-superactivation requires only the adding of *quantum entanglement* and the use of *stimulated emission* [29] [36]; then *arbitrary* zero-capacity quantum channels can be used for classical communication. While the superactivation of classical capacity of quantum channels is trivially not possible, here we prove that the quasi-superactivation is possible, and information can be transmitted by the addition of noise to the quantum channel. We show that classical information can also be transmitted over the combination of zero-capacity quantum channels using *quasi-superactivation* and it requires only the most natural process that occurs during stimulated emission. Another important difference that while the superactivation of classical capacity of quantum channels *is theoretically not possible*, here we prove that the quasi-superactivation *is possible*.

Let's describe in detail the difference between our proposal of quasi-superactivation and the original superactivation [6-8] [13] [15]. The superactivation of classical capacity of two zero classical-capacity quantum channels is not possible by the following simple reason. The superactivation of zero-capacity quantum channels is based on a *conversion* between the channel capacities: for example in the case of superactivation of quantum capacity [13-14] the private classical capacity of one of the channels from the joint channel construction was "converted" to a smaller amount of quantum capacity, which was obtained by the use of the joint channel structure (*Note*: in the proof of Smith and Yard's [13] this channel was the so-called Horodecki channel [40]). If the two channels are not capable of transmitting classical information (i.e., they cannot preserve any classical correlation), then these channels trivially have no remain other "convertible" channel capacities which would be a required initial condition for the superactivation of channel capacities. (*Note*: The impossibility of superactivation of classical capacity also can be proven mathematically by the *Choi-Jamiolkowski isomorphism* [37-38], - we omit the proof here.) In the proposed quasi-superactivation effect we have a completely different scenario in comparison to the originally introduced superactivation effect. We do not require the existence of any initial positive classical capacity to achieve positive joint channel capacity, since quasi-superactivation is based on a different procedure than the previously mentioned *capacity conversion*. From this difference an important conclusion follows: while the originally introduced superactivation cannot be extended for the classical capacity, the quasi-superactivation can, because this effect is based on a different assumption. Roughly speaking, in the case of quasi-superactivation we will extract valuable information from an auxiliary input system which initially was completely independent from the original input system. The quasi-superactivation does not require any *capacity conversion*, which makes it possible to extend it to the case of classical capacity. As a consequence, positive classical correlation can be achieved between the output of the joint channel of zero classical-capacity channels and the classical register.

This paper is organized as follows. In Section 2 we summarize the important functions and definitions regarding the background of the proof. In Section 3 we describe the proposed channel setting. In Section 4 we present the theorems and proofs. Section 5 demonstrates the results, finally in Section 6 we conclude the paper. Further supplementary information is included in the Appendix.

# 2 Preliminaries

Quantum channels extend the possibilities, and besides the classical information we can send entanglement-assisted classical information, classical private information, and of course, quantum information. On the other hand, the elements of classical information theory cannot be applied in general for quantum information—in other words, they can be used only in some special cases. There is no general formula to describe the capacity of every quantum channels, but one of the main recent results in quantum Shannon theory is that many of quantum channel capacities are in fact non-additive.

Quantum Information Processing exploits the quantum nature of information. It offers fundamentally new solutions in the field of computer science and extends the possibilities to a level that cannot be imagined in classical communication systems. On the other hand, it requires the generalization of classical information theory through a quantum perception of the world. Thanks to Shannon we can calculate the capacity of classical channels within the frames of classical information theory. However, in order to measure the maximum amount of classical information that can be sent over quantum channels, we have to redefine the well-known formulas of classical information theory. In case of a quantum communication system, the information sent through quantum channels is carried by quantum states, hence the encoding is fundamentally different from any classical encoder scheme. The encoding here means the preparation of a quantum system, according to the probability distribution of the classical message being encoded. Similarly, the decoding process is also different: here it means the measurement of the received quantum state. These fundamental differences between the classical and quantum systems cannot be described without the elements of Quantum Information Theory.

## 2.1 Quantum Entropy and the Holevo Information

As Shannon entropy plays a fundamental role in classical information theory, the von Neumann entropy does the same for quantum information. The von Neumann entropy $S(\rho)$ of quantum state $\rho$ can be viewed as an extension of classical entropy for quantum systems. It measures the information of the quantum states in the form of the uncertainty of a quantum state [15] [20]. The classical Shannon entropy $H(X)$ of a variable $X$ with probability distribution $p(X)$ can be defined as $H(X) = -\sum_{x \in X} p(x) \log(p(x))$, with $1 \leq H(X) \leq \log(|X|)$, where $|X|$ is the cardinality of the set $X$. The von Neumann entropy $S(\rho) = -Tr(\rho \log(\rho))$ measures the information contained in the quantum system $\rho$. Furthermore $S(\rho)$ can be expressed by means of the Shannon entropy for the eigenvalue distribution $S(\rho) = H(\lambda) = -\sum_{i=1}^{d} \lambda_i \log(\lambda_i)$, where $d$ is the level of the quantum system and $\lambda_i$ are the eigenvalues of density matrix $\rho$.

The *Holevo bound* determines the classical information that can be extracted from a quantum state. If Alice sends a quantum state $\rho_i$ with probability $p_i$ over an ideal quantum channel, then at Bob's receiver a mixed state $\rho_B = \rho_A = \sum_i p_i \rho_i$ appears. Bob constructs a measurement $\{M_i\}$ to extract the information encoded in the quantum states. If he applies the measurement to $\rho_A$, the probability distribution of Bob's classical symbol $B$ will be $\Pr[b|\rho_A] = Tr(M_b^\dagger M_b \rho_A)$. As had been shown by Holevo [10], the bound for the maximal classical mutual information between Alice and Bob is

$$I(A:B) \leq S(\rho_A) - \sum_i p_i S(\rho_i) \equiv \chi, \tag{1}$$

where $\chi$ is called the *Holevo quantity*. In classical information theory and classical communication systems, the mutual information $I(A:B)$ is bounded only by the classical entropy of $H(A)$, hence $I(A:B) \leq H(A)$.

## 2.2 Classical Capacity of a Quantum Channel

The *classical capacity* $C(\mathcal{N})$ of a quantum channel describes the amount of classical information that can be transmitted through the channel. One of the earliest works on the capacities of quantum communication channels was published in the early 1970s [10]. Along with other researchers, Holevo showed that there are many differences between the properties of classical and quantum communication channels, and illustrated this with the benefits of using entangled input states. After Holevo published his work [10], about 30 years later he, with Schumacher and Westmoreland, presented one of the most important results in Quantum Information Theory, called the *Holevo-Schumacher-Westmoreland* (HSW) theorem [11-12]. The HSW theorem defines the maximum amount of classical information that can be transmitted through a noisy quantum channel $\mathcal{N}$ if the input contains product states (i.e., entanglement is not allowed). In this setting, for the noisy quantum channel $\mathcal{N}$ the $C(\mathcal{N})$ classical capacity (which is also referred in the literature as the $\chi(\mathcal{N})$ Holevo capacity of a quantum channel) can be expressed as

$$\begin{aligned} C(\mathcal{N}) &= \chi(\mathcal{N}) \\ &= \max_{\forall\, p_i, \rho_i} \chi = \max_{\forall\, p_i, \rho_i} \left[ S(\sigma_{out}) - \sum_i p_i S(\sigma_i) \right] \\ &= \max_{\forall\, p_i, \rho_i} \left[ S\left( \mathcal{N}\left( \sum_i p_i \rho_i \right) \right) - \sum_i p_i S(\mathcal{N}(\rho_i)) \right], \end{aligned} \qquad (2)$$

where the maximum is taken over all ensembles $\{p_i, \rho_i\}$ of input quantum states, while $\sigma_{out} = \mathcal{N}\left( \sum_i p_i \rho_i \right)$ is the average channel output and $\mathcal{N}(\rho_i)$ are the channel output states. The HSW theorem is a generalization of the classical noisy channel-coding theorem from classical information theory to a noisy quantum channel. On the other hand, the HSW theorem also raised a lot of questions regarding the transmission of classical information over general quantum channels. Hastings showed that the entangled inputs can increase the amount of classical information [9], for instance, $\chi(\mathcal{N}) = \max_{\forall\, p_i, \rho_i} \chi \neq C(\mathcal{N})$ and the $C(\mathcal{N})$ can be expressed by the asymptotic formula of Holevo capacity $\chi(\mathcal{N})$ as

$$C(\mathcal{N}) = \lim_{n \to \infty} \frac{1}{n} \chi(\mathcal{N}^{\otimes n}), \qquad (3)$$

where $\mathcal{N}^{\otimes n}$ denotes the *n* uses of the quantum channel $\mathcal{N}$.

## 2.3 The Cloning Quantum Channel

One of the key components for the security of QKD (*Quantum Key Distribution*) is the impossibility of cloning of an unknown quantum state [17]. Contrary to classical information, in a quantum communication system the quantum information cannot be copied perfectly. If Alice sends a number of photons $|\psi_1\rangle, |\psi_2\rangle, ..., |\psi_N\rangle$ through the quantum channel and if Eve wants to copy the *i*-th sent photon $|\psi_i\rangle$, she has to apply a unitary transformation $U$, which gives the following result:

$$U(|\psi_i\rangle \otimes |0\rangle) = |\psi_i\rangle \otimes |\psi_i\rangle. \tag{4}$$

A photon chosen from a given set of polarization states can be cloned perfectly only if the polarization angles in the set are all mutually orthogonal. Unknown non-orthogonal states cannot be cloned perfectly, and the cloning process of the quantum states is possible only if the information being cloned is classical. The cloning channel [1] is based on the *Universal Quantum Cloner Machine* (UQCM), which was defied in 1996 [3]. The output fidelity of the cloned state is independent from the fidelity of the input system. The process of cloning of pure states can be generalized as

$$|\psi\rangle_a \otimes |\Sigma\rangle_b \otimes |Q\rangle_x \rightarrow |\Psi\rangle_{abx}, \tag{5}$$

where $|\psi\rangle$ is the state in the Hilbert space to be copied, $|\Sigma\rangle$ is a reference state, and $|Q\rangle$ is the ancilla system [3-4]. A cloning machine is considered symmetric if at the outputs, all the clones have the same fidelity, and asymmetric if the clones have different fidelities.

In the case of the quantum-cloning channel, the fidelity of the channel output state $\rho$ is independent of the input quantum state $|\psi\rangle$, and the fidelity of the output system can be expressed as $F = \langle\psi|\rho|\psi\rangle$. In general, for a universal quantum cloner

$$F = \left(\frac{2}{3}\right) + \left(\frac{1}{3N}\right), \tag{6}$$

where *N* is the number of channel output states.

As follows from (6), the fidelity of the cloning process depends only on the number $N$ of produced output states. The largest classical capacity of a cloning quantum channel $\mathcal{N}_2$ can be achieved if the channel realizes an $1 \rightarrow 2$ cloning process, i.e. the number of cloned output states is $N - 1 = 2 - 1 = 1$ [1-2] [4].

Assuming two output systems $\rho_1$ and $\rho_2$ of the cloning quantum channel, the channel generates outputs with the same properties independently from the input of the channel, i.e., $\rho_1 = \rho_2$. The fact that every $1 \rightarrow N$ quantum-cloning channel is *entanglement breaking* (a quantum channel is called entanglement breaking if it destroys every entanglement on the outputs and produces separable output states whenever the input of the channel is an entangled system) was shown in 2009 [1]. We note that this fact does not contradict to our aims and we can apply these types of quantum channels in the quasi-superactivation of classical capacity.

## 2.4 Channel System Description

Here we summarize and review the basic elements of the proposed system model. The system model consists of the following elements: Alice's classical register $X$, the purification state $P$, channel input $A$, channel output $O$, and the environment state $E$. The input system $A$ is described by a quantum system $\rho_x$, which occurs on the input with probability $p_X(x)$ [16]. They together form an ensemble denoted by $\{p_X(x), \rho_x\}_{x \in X}$, where $x$ is a classical variable from the classical register $X$. In the preparation process, Alice generates pure states $\rho_x$ according to random variable $x$, i.e., the input density operator can be expressed as $\rho_x = |x\rangle\langle x|$, where the classical states $\{|x\rangle\}_{x \in X}$ form an orthonormal basis [20]. According to the elements of Alice's classical register $X$, the input system can be characterized by the quantum system

$$\rho_A = \sum_{x \in X} p_X(x) \rho_x = \sum_{x \in X} p_X(x) |x\rangle\langle x|.$$

The system description is illustrated in Fig. 1.

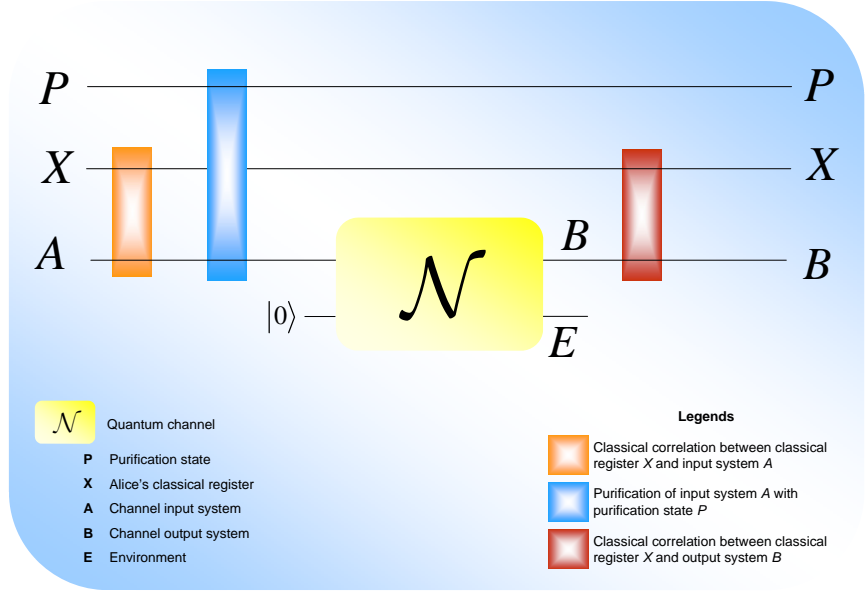

**Fig. 1.** Alice's classical register is denoted by X, the input system is A while P is the purification state. The environment of the channel is denoted by E, the output of the channel is B. The quantum channel has positive classical capacity if and only if the channel output system B will be correlated with Alice's classical register X.

The system state $\rho_x$ with the corresponding probability distribution $p_X(x)$ can be indentified by a set of measurement operators $M = \{|x\rangle\langle x|\}_{x \in X}$. If the density operators $\rho_x$ in $\rho_A$ are mixed, the probability distribution $p_X(x)$ and the classical variable $x$ from the register $X$ cannot be indentified by the measurement operators $M = \{|x\rangle\langle x|\}_{x \in X}$, since the system state $\rho_x$ is assumed to be a mixed or in a non-orthonormal state.

Alice's classical register $X$ and the quantum system $A$ can be viewed as a tensor product system as $\{p_X(x), |x\rangle\langle x|_X \otimes \rho_A^x\}_{x \in X}$, where the quantum state $|x\rangle$ is correlated with the quantum system $\rho_x$, using orthonormal basis $\{|x\rangle\}_{x \in X}$. Alice's register $X$ represents a classical variable, the channel input system is generated corresponding to the register $X$ in the form of a quantum state, and it is described by the density operator $\rho_A^x$. The input system $A$ with to the classical register $X$, is described by the density operator $\rho_{XA} = \sum_{x \in X} p_X(x) |x\rangle\langle x|_X \otimes \rho_A^x$, where $\rho_A^x = |\psi_x\rangle\langle\psi_x|_A$.

# 3 Classical Communication over Zero-Capacity Links

The superactivation of zero-capacity quantum channels makes it possible to use two zero-capacity quantum channels with a positive joint capacity for their output. Currently, we have no theoretical background to describe all possible combinations of superactive zero-capacity channels; hence, there may be many other possible combinations [21]. In 2008, Smith and Yard have found only one possible combination for superactivation of quantum capacity [13], and later it was extended to the classical zero-error [7-8] and quantum zero-error capacities of quantum channels [6]. While the superactivation of classical capacity is not possible, as we have proven, the quasi-superactivation of classical capacity is possible. Moreover, as we have found, it works for the most generalized quantum channel models, which describe the most natural physical processes.

In Fig. 2, we show our channel construction $\mathcal{M} = \mathcal{N}_1 \circ \mathcal{N}_2$, where $\circ$ represents the channel concatenation. The first channel $\mathcal{N}_1$ can be any zero-capacity quantum channel that produces a maximally mixed output state (The channel output is completely uncorrelated with the input, i.e., the classical capacity of the channel is zero since the channel destroys every classical correlation.). The second channel $\mathcal{N}_2$ in the channel construction $\mathcal{M}$ is the so-called cloning quantum channel [1-2], which channel model describes a natural process that occurs during *stimulated emission* [28-29].

The cloning channel [1-2] describes the effect of optical amplification as a result of the fundamental interaction of an atom with an impinging photon. The effect is known as stimulated emission and occurs, for instance, in erbium-doped optical fibers [29-31] [35]. Furthermore, it was also found that the qudit Unruh channel has deep connection with the cloning channels [28] [39].

As follows the stimulated emission has deep relevance in the quasi-superactivation of the classical capacity of quantum channels. The process that occurs during the stimulated emission is described and modeled by the cloning quantum channel $\mathcal{N}_2$ [1] [28-29].

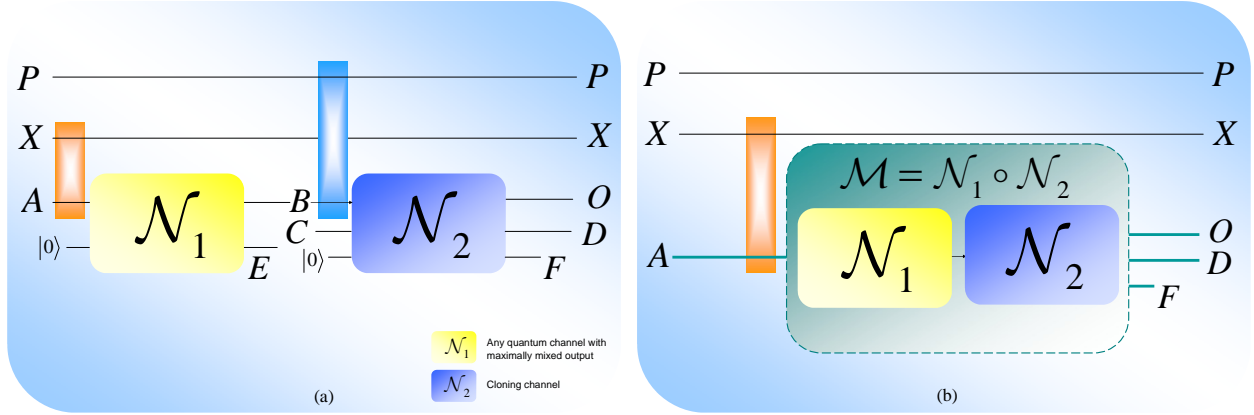

**Fig. 2.** (a): The first quantum channel $\mathcal{N}_1$ can be any quantum channel that produces a maximally mixed output state. The second channel $\mathcal{N}_2$ is the cloning channel. Alice's classical register is denoted by $X$, the channel input is $A$, while $P$ is the purification state (it describes the connection with the environment). In the sending process Alice correlates her quantum system $A$ with her classical register $X$ (orange-shaded rectangle). The first channel destroys every classical correlation between register $X$ and channel output $B$. The input of the second channel is the output $B$ of the first channel. The purification of system $B$ is denoted by the blue-shaded rectangle. The environment of the first channel is depicted by $E$. The output of the channel is $O$, while $D$ is the cloned output and $F$ is the environment. (The environments of the channels are initialized in the pure input system $|0\rangle$.)

(b): The quantum channel $\mathcal{M}$ cannot transmit any classical information. It consists of channels $\mathcal{N}_1 \circ \mathcal{N}_2$ where $\mathcal{N}_1$ can be any quantum channel that generates a maximally mixed output state and $\mathcal{N}_2$ is the cloning quantum channel, the classical capacity is zero, i.e., $C(\mathcal{M}) = 0$. (For the proof see Section 4.)

As we will prove, while individually quantum channel $\mathcal{M}$ cannot used to transmit any classical information (Fig. 3(a)), but something strange thing will occur if we use together two of these zero-capacity channels $\mathcal{M}_1$ and $\mathcal{M}_2$, in which each channel $\mathcal{M}_i$, $i = 1,2$ is constructed from a zero-capacity quantum channel $\mathcal{N}_1$ and an $1 \to N$ cloning quantum channel (Fig. 3(b)). As we have found, two zero-capacity quantum channels in a joint structure $\mathcal{M}_1 \otimes \mathcal{M}_2$ can activate each other, and the joint classical capacity will be positive, while for the individual classical capacities $C(\mathcal{M}_1) = C(\mathcal{M}_2) = 0$.

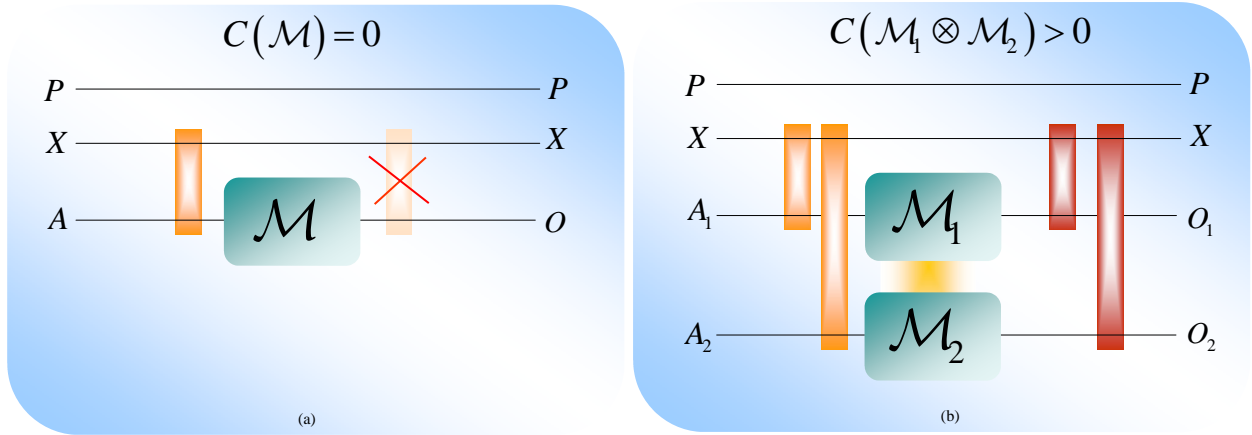

**Fig. 3.** (a): Individually quantum channel $\mathcal{M}$ cannot transmit any classical information, i.e., $C(\mathcal{M}) = 0$. The channel destroys every classical correlation between Alice's classical register $X$ and channel output $O$.

(b): For the joint combination of the two zero-capacity quantum channels $\mathcal{M}_1$ and $\mathcal{M}_2$ with classical capacities $C(\mathcal{M}_1) = C(\mathcal{M}_2) = 0$, the joint classical capacity will be positive, i.e., $C(\mathcal{M}_1 \otimes \mathcal{M}_2) > 0$. Any correlation between classical register $X$ and output systems $O_1$ and $O_2$ will occur that result in positive classical capacity (For the proof see Section 4.).

The no-cloning theorem [17] is one of the most important fundaments of Quantum Information Theory, since it makes impossible to copy the (unknown, non orthogonal) quantum states—which is a trivial process in classical systems, where the classical bits can be copied arbitrary many times. In the case of a quantum system, the picture changes completely. An arbitrary quantum system cannot be copied perfectly; however, with lower output fidelity the cloning can be realized. This was an important discovery in 1996 [3] in QIT, since before it was thought (and a generally accepted fact) that the no-cloning theorem forbids any possible cloning process on a quantum system [17]. The definition of *cloning quantum channel* first appeared in Quantum Information Theory just in 2009 [1], although the theoretical basis was studied before that [3-5], [29-34]. The cloning quantum channel makes it possible to generate copied quantum states at the channel output without violating the no-cloning theorem. The output of a cloning channel consists of the original input qubit and $N$-1 copied states; this quantum channel is called the $1 \to N$ cloning channel. The fidelity of the cloning process and the classical capacity of the cloning channel decreases as the number of $N$ increases. This decrease in the capacity is the price we have to pay to avoid violating the no-cloning theorem [1-4]. As we derived in our proof, the quasi-superactivation of classical capacity requires special conditions in the input system. (For the further mathematical background, see References [18-19]). As we have found, the cloning channels can be very useful in the quasi-superactivation of classical capacity. In the proof we also reveal that the channel construction $\mathcal{M}_1 \otimes \mathcal{M}_2$ of zero-capacity channels can be used for transmission of classical information only in a very small parameter domain.

Fig. 4 helps explain what is happening in the background and how the quasi-superactivation of classical capacity works.

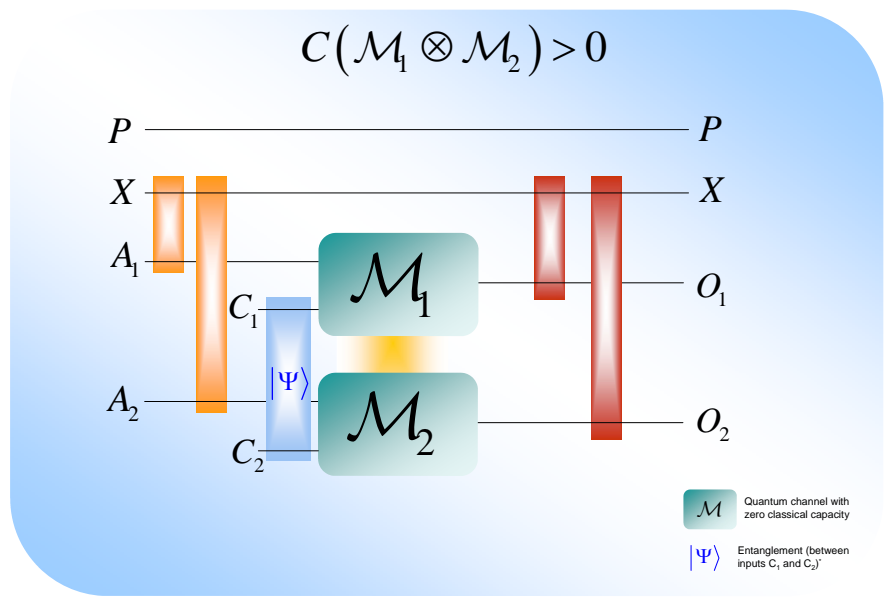

**Fig. 4.** The addition of *noise* in the form of entanglement. The detailed view of joint channel construction $\mathcal{M}_1 \otimes \mathcal{M}_2$ helps to reveal the effect. Individually, neither $\mathcal{M}_1$, nor $\mathcal{M}_2$ can transmit any classical information. On the other hand, if we use entangled auxiliary input and the amount of entanglement in the input qubits is chosen from a very limited domain, the two channels can activate each other and classical information can be transmitted. Using input states with this special amount of entanglement, the outputs of the joint channel construction will be correlated with each other and some correlation will also occur with the classical register *X*. However, individually every classical correlation will vanish, jointly some correlation can be produced at the channel output which leads to positive classical capacity.

The theorems and proofs regarding the quasi-superactivation of quantum channels will be presented in Section 4.

# 4 Theorems and Proofs

In this section we give the mathematical proof of quasi-superactivation of classical capacity.

## 4.1 Proof of Quasi-Superactivation of Classical Capacity

Here give the mathematical proof of quasi-superactivation of the classical capacity of the joint structure $\mathcal{M} = \mathcal{N}_1 \circ \mathcal{N}_2$.

The input system of the channel is described as follows. Alice correlates her classical register $X$ with her pure quantum system $\rho_A^x = |\psi_x\rangle\langle\psi_x|_A$ and then puts to the input $A$ of the first channel. The environment $E$ is also initialized in the pure state $|0\rangle$. The output of the first channel is $B$. Alice's input system is $|\psi_x\rangle_A \otimes |0\rangle_E$, which will be used in the description of the evolution of the channel, using the unitary transformation $U_{AE}$. The transformation $U_{AE}$ describes the interaction between the input system $A$ and the environment $E$. Using $|\psi_x\rangle_A \otimes |0\rangle_E$ the unitary interaction is $U_{AE}(|\psi_x\rangle_A \otimes |0\rangle_E)$. (For further information see Appendix.)

The purification is an abstract picture, which describes the noise of the quantum channel in terms of entanglement between an inaccessible purification state $P$ and the (generally) mixed channel output state $\sigma_B$ [16] [20]. The purification state $P$ also can be viewed as the external environment of the quantum channel $\mathcal{N}$, and according to the interaction of the channel the quantum state and the environment, becomes entangled. (The environment $E$ is the purification of the noisy output system $B$ of the quantum channel.) The interaction between the quantum system $\rho$ and the environment $P$ causes information loss, since the purification state $P$ is not accessible.

In our channel construction $\mathcal{M} = \mathcal{N}_1 \circ \mathcal{N}_2$, the purification also has a very useful application, since it can help us to understand the working process of the first, completely noisy quantum channel $\mathcal{N}_1$ and to describe the input system of the cloning quantum channel $\mathcal{N}_2$. The first channel $\mathcal{N}_1$ can be any quantum channel that produces a maximally mixed state. See the output of the first channel $\mathcal{N}_1$. Assuming a qubit channel, i.e., $d=2$, it is the maximally mixed state $\sigma_B = \frac{1}{2}I = \frac{1}{2}(|0\rangle\langle 0| + |1\rangle\langle 1|)$. The purification of the maximally mixed system $B$ is the Bell state

$$|\varphi\rangle_{PB} = |\Phi_{00}\rangle = \frac{1}{\sqrt{2}}(|00\rangle + |11\rangle), \tag{7}$$

where $P$ is the purification state. As follows, after Alice has transmitted her quantum system $\rho_A$ over the noisy quantum channel $\mathcal{N}_1$, the system becomes a completely mixed state, which forms the maximally entangled state $|\Phi_{00}\rangle$ with the purification state $P$ [1-2].

Assuming probability amplitudes $\alpha$ and $\beta$ the $PB$ purified system state

$$|\varphi\rangle_{PB} = \alpha|00\rangle + \beta|11\rangle \tag{8}$$

of the input of the second channel can be expressed as:

$$|\varphi\rangle\langle\varphi|_{PB} = \alpha^2|00\rangle\langle 00| + \alpha\beta|00\rangle\langle 11| + \beta\alpha|11\rangle\langle 00| + \beta^2|11\rangle\langle 11|, \tag{9}$$

where $\alpha^2 \leq 0.5$ and $\beta^2 = 1 - \alpha^2$ denote the probabilities (square of the probability amplitudes) and $\alpha^2 + \beta^2 = 1$.

The transformation of the cloning channel $\mathcal{N}_2$ can be characterized as follows (see Fig. 2). The first input $(B)$ of the channel is the original quantum state to be cloned. The second input of the channel $(C)$ is a blank state initialized with $|0\rangle$. The environment $F$ of the second channel is also initialized with $|0\rangle$. Generally, the input $C$ can be any state, the cloning channel can be described without any characterization of this state—meaning its specification can be omitted from the system description. (We note that in the quasi-superactivation of the classical capacity of the joint channel structure $\mathcal{M}_1 \otimes \mathcal{M}_2$ this input will get an entangled system, i.e., it will have deep relevance in the quasi-superactivation process.) We also do not show the initial and the final state of the cloning channel, since they no relevance in the description. The final state of the universal cloner machine can be described by two orthogonal vectors in the Hilbert space, according to the two possible outcomes.

Assuming a $1 \to 2$ cloning channel $\mathcal{N}_2$ (which will be also used in our construction, since the best classical capacity can be reached with this cloning channel, because the classical capacity of the cloning channel decreases as $N$ increases [1-2]), the working mechanism of the probabilistic cloning channel can be expressed as follows:

$$
\begin{aligned}
|0\rangle_B |0\rangle_C &\xrightarrow{\mathcal{N}_2} \sqrt{\tfrac{2}{3}}|00\rangle_{OD} + \sqrt{\tfrac{1}{3}}\left(\tfrac{1}{\sqrt{2}}(|01\rangle + |10\rangle)\right)_{OD}, \\
|1\rangle_B |0\rangle_C &\xrightarrow{\mathcal{N}_2} \sqrt{\tfrac{2}{3}}|11\rangle_{OD} + \sqrt{\tfrac{1}{3}}\left(\tfrac{1}{\sqrt{2}}(|01\rangle + |10\rangle)\right)_{OD},
\end{aligned}
\tag{10}
$$

where $B$ and $C$ are the inputs, $O$ and $D$ are the outputs. For an unknown, un-entangled pure input quantum system $|\psi\rangle = \alpha|0\rangle + \beta|1\rangle$ the output density matrices $\rho_O$ and $\rho_D$ are equal, i.e.,

$$
\rho_O = \rho_D = \tfrac{5}{6}|\psi\rangle\langle\psi| + \tfrac{1}{6}|\psi^\perp\rangle\langle\psi^\perp|.
\tag{11}
$$

This result means that the $1 \to 2$ cloning channel $\mathcal{N}_2$ with probability $5/6$ puts the outputs to $OD$ the original input system, otherwise, with probability $1/6$ it puts the orthogonal state to the original $|\psi\rangle$ to outputs $OD$ of the channel $\mathcal{M}$.

The input of the $1 \to 2$ cloning channel $\mathcal{N}_2$ is depicted by $B$. The two outputs of the cloning channel are depicted by $O$ and $D$ (the two outputs together will be denoted by $OD$), while for the third output $F$ (the environment state), see Fig. 2. Assuming an $1 \to N$ cloning channel, the channel transformation on an unknown quantum system $|\psi\rangle$ can be expressed as (omitting the $N$ completely symmetric orthonormal basis states $|N-1, i\rangle$ of the channel from the description)

$$
\begin{aligned}
|0\rangle_B |\psi\rangle_C &\xrightarrow{\mathcal{N}_2} \sum_{i=0}^{N-1} \tau_i^{(N)} |N, i\rangle_{OD}, \\
|1\rangle_B |\psi\rangle_C &\xrightarrow{\mathcal{N}_2} \sum_{i=0}^{N-1} \tau_{N-1-i}^{(N)} |N, i+1\rangle_{OD},
\end{aligned}
\tag{12}
$$

where

$$
\tau_i^{(N)} = \sqrt{2\frac{(N-i)}{N(N+1)}}.
\tag{13}
$$

The $U$ unitary map (see Isometric Extension in Appendix) of the cloning channel $\mathcal{N}_2$ on a given input system $\rho$ can be described as

$$\mathcal{N}_2\left(U\rho U\right) = W\mathcal{N}_2\left(\rho\right)W^\dagger, \qquad (14)$$

where $\mathcal{N}_2\left(\rho\right)$ is the output state and $W$ is the irreducible representation [2] [26] of unitary transformation $U$.

To describe the working mechanism of the cloning quantum channel we expressed it in the form of *isometric extension*. For simplicity, the input system $B$ is described on the $\{|0\rangle,|1\rangle\}$ basis, while the output $OD$ is written as

$$\{|b\rangle_{OD} = |N-b,b\rangle\} = \{|b\rangle_{OD} = |2-b,b\rangle\},\ for\ 0 \leq b \leq N = 2, \qquad (15)$$

where $|b\rangle_{OD}$ is an orthonormal basis of the output system $OD$, which consists of completely symmetric states (since the cloning channel produces both output qubits with the same fidelity), and $|2-b,b\rangle_{OD}$, for $0 \leq b \leq N = 2$ is a normalized state of the two-qubit output system $OD$ of the cloning channel $\mathcal{N}_2$, which represents the superposition of the two-qubit basis states

$$\{|00\rangle,|01\rangle,|10\rangle,|11\rangle\}_{OD}. \qquad (16)$$

Now, we describe from the environment $F$ point of view the output of the zero-capacity quantum channel $\mathcal{M}$. The orthonormal basis for the environment $F$ is given as

$$\{|k\rangle_F = |N-k-1,k\rangle\} = \{|k\rangle_F = |2-k-1,k\rangle\},\ for\ 0 \leq k \leq N-1 = 2-1, \qquad (17)$$

where $|2-k,k\rangle_F$, for $0 \leq k \leq 2-1$ is the normalized state on the $2-1=1$ quantum state. The environment $F$ is in a superposition of the basis states $|k\rangle_F \in \{|0\rangle,|1\rangle\}_F$.

The isometric extension $U_{\mathcal{N}_2}^{B\rightarrow(OD)F}$ of the cloning channel $\mathcal{N}_2$ from the channel structure $\mathcal{M}$ for an $1 \rightarrow N$ and $1 \rightarrow 2$ channel is [1]

$$U_{\mathcal{N}_2}^{B\to(OD)F} = \frac{1}{\sqrt{\Delta_N}}\sum_{k=0}^{N-1}\sqrt{N-1}|k\rangle_{OD}\langle 0|_B \otimes |k\rangle_F + \frac{1}{\sqrt{\Delta_N}}\sum_{k=0}^{N-1}\sqrt{k+1}|k+1\rangle_{OD}\langle 1|_B \otimes |k\rangle_F$$

$$= \frac{1}{\sqrt{\Delta_2}}\sum_{k=0}^{1}|k\rangle_{OD}\langle 0|_B \otimes |k\rangle_F + \frac{1}{\sqrt{\Delta_2}}\sum_{k=0}^{1}\sqrt{k+1}|k+1\rangle_{OD}\langle 1|_B \otimes |k\rangle_F,$$

(18)

where

$$\Delta_N = \frac{N(N+1)}{2}.$$

(19)

The Kraus representation of the cloning quantum channel $1 \to N$ (and $1 \to 2$) can be expressed as

$$\mathcal{N}_2 = \frac{1}{\sqrt{\Delta_N}}\left(\sqrt{N-k}|k\rangle_{OD}\langle 0|_B + \sqrt{k+1}|k+1\rangle_{OD}\langle 1|_B\right) \text{ for } 0 \le k \le N\text{-}1,$$

$$\mathcal{N}_2 \xrightarrow{1\to 2} \left\{\frac{1}{\sqrt{\Delta_2}}\left(\sqrt{2}|0\rangle_{OD}\langle 0|_B + |1\rangle_{OD}\langle 1|_B\right), \frac{1}{\sqrt{\Delta_2}}\left(|1\rangle_{OD}\langle 0|_B + \sqrt{2}|2\rangle_{OD}\langle 1|_B\right)\right\},$$

(20)

while the K Kraus operators for the complementary channel of the cloning quantum channel are

$$\mathrm{K} = \left\{\sqrt{N}|0\rangle_F\langle 0|_B, \sqrt{N-k}|k\rangle_F\langle 0|_B + \sqrt{k}|k-1\rangle_F\langle 1|_B, \sqrt{N}|N-1\rangle_F\langle 1|_B, \text{ for } 1 \le k \le N-1\right\}$$

$$\xrightarrow{1\to 2} \left\{\sqrt{2}|0\rangle_F\langle 0|_B, |1\rangle_F\langle 0|_B + |0\rangle_F\langle 1|_B, \sqrt{2}|1\rangle_F\langle 1|_B\right\}.$$

(21)

For the $1 \to 2$ cloning channel $\mathcal{N}_2$ these Kraus operators can be rewritten in the following way:

$$\mathrm{K} = \left\{\sqrt{\frac{1}{3}}|e_1(X)\rangle\langle e_1(X)|, \sqrt{\frac{1}{3}}|e_2(X)\rangle\langle e_2(X)|,\right.$$
$$\left.\sqrt{\frac{1}{3}}|0\rangle\langle 0|, \sqrt{\frac{1}{3}}|1\rangle\langle 1|, \sqrt{\frac{1}{3}}|e_1(Y)\rangle\langle e_1(Y)|Z, \sqrt{\frac{1}{3}}|e_2(Y)\rangle\langle e_2(Y)|Z\right\},$$

(22)

where $\{e_1(X), e_2(X)\}$ are the eigenvalues of the Pauli $X$ matrix, $\{e_1(Y), e_2(Y)\}$ are the eigenvalues of the Pauli $Y$ matrix, and $Z$ is the Pauli $Z$ transformation [2].

**Theorem 1.** *For a quantum channel $\mathcal{M} = \mathcal{N}_1 \circ \mathcal{N}_2$, where $\mathcal{N}_1$ can be any quantum channel that generates maximally mixed output state and $\mathcal{N}_2$ is the cloning quantum channel, the classical capacity is zero, i.e., $C(\mathcal{M}) = 0$.*

First, we prove that $\mathcal{N}_1$ has zero classical capacity. In the second part, we show that the cloning channel $\mathcal{N}_2$ also has zero classical capacity.

*First Part of Proof of Theorem 1.*

Here we prove that quantum channel $\mathcal{M}$ has zero classical capacity, i.e., it cannot be used to transmit any classical information.

In the first part of the proof we show that the classical capacity of $\mathcal{N}_1$ is zero. The first channel $\mathcal{N}_1$ is a quantum channel that generates a maximally mixed output state $\sigma_{out} = \frac{1}{d} I_d$, where $I_d$ is the $d$ dimensional identity matrix. (Assuming $d=2$, $\sigma_{out} = \frac{1}{2} I$.) The von Neumann entropy of this state is $S(\sigma_{out}) = 1$ and, according to the Holevo-Schumacher-Westmoreland theorem [11-12] the classical capacity can be expressed as

$$C(\mathcal{N}_1) = \chi(\mathcal{N}_1) = \max_{\forall\, p_i, \rho_i} \chi = \max_{\forall\, p_i, \rho_i} \left[ S(\sigma_{out}) - \sum_i p_i S(\sigma_i) \right] =$$
$$= \max_{\forall\, p_i, \rho_i} \left[ S\left(\mathcal{N}_1\left(\sum_i p_i \rho_i\right)\right) - \sum_i p_i S(\mathcal{N}_1(\rho_i)) \right] \qquad (23)$$
$$= 1 - 1 = 0,$$

where $\chi$ is the Holevo quantity defined as $\chi = S\left(\mathcal{N}_1\left(\sum_i p_i \rho_i\right)\right) - \sum_i p_i S(\mathcal{N}_1(\rho_i))$.

*Illustration of the Proof*

For the quantum channel $\mathcal{N}_1$ (for example it can be a completely depolarizing channel [16]) the classical capacity is trivially zero, since in (23)

$$S\left(\mathcal{N}_1\left(\sum_i p_i \rho_i\right)\right) = \sum_i p_i S(\mathcal{N}_1(\rho_i)) = 1, \qquad (24)$$

where $\mathcal{N}\left[\sum_i p_i \rho_i\right]$ is the average output state of the channel $\mathcal{N}_1$. In conclusion, any channel $\mathcal{N}_1$ that produces a maximally mixed output destroys every classical correlation between Alice's register $X$ and the output $B$ of the channel, and its classical capacity is $C(\mathcal{N}_1) = 0$.

To illustrate the result of (23), let us assume $\mathcal{N}_1$ is a completely depolarizing qudit ($d>2$ dimensional) channel that generates a maximally mixed state for its every input. The map of the completely depolarizing channel can be given as

$$\mathcal{N}_{Compl.Depol.}(\rho_A) = (1-p)\rho_A + p\left(\frac{1}{d}I_d\right), \qquad (25)$$

where $p = 1$, and $\frac{1}{d}I_d$ is a maximally mixed one-qudit state [20] (If we have a qubit channel, then $d=2$, i.e., $\frac{1}{2}I$.).

**Lemma 1.1.** The classical capacity of $\mathcal{N}_1 = \mathcal{N}_{Compl.Depol.}$, assuming a qudit channel with arbitrary dimension $d$ (if $d = 2$, then we talk about a qubit channel) is

$$\begin{aligned} C(\mathcal{N}_1) &= \log_2(d) + \left[1 - p + \frac{p}{d}\right]\log_2\left[1 - p + \frac{p}{d}\right] + (d-1)\frac{p}{d}\log_2\left(\frac{p}{d}\right) \\ &= \log_2(d) + \frac{1}{d}\log_2\left(\frac{1}{d}\right) + (d-1)\frac{1}{d}\log_2\left(\frac{1}{d}\right) = 0. \end{aligned} \qquad (26)$$

For example for the qubit $(d = 2)$ case

$$C(\mathcal{N}_1) = \log_2(2) + \frac{1}{2}\log_2\left(\frac{1}{2}\right) + \frac{1}{2}\log_2\left(\frac{1}{2}\right) = 0. \qquad (27)$$

*Proof of Lemma 1.1.*

Here we prove that the Holevo information of the completely depolarizing channel $\mathcal{N}_1$ is zero.

In the first step, we show that the minimum output entropy of the completely depolarizing channel $\mathcal{N}_1$ is 1, i.e., $S_{\min}(\mathcal{N}_1(\rho_A)) = 1$, or with other words we prove that the output of the channel is a maximally mixed state, independent from the channel input. Here we assume that the minimum output entropy $S_{\min}(\mathcal{N}_1(\rho_A))$ of the channel $\mathcal{N}_1$ is the minimum of the entropy taken at the output of the channel $\mathcal{N}_1$, i.e., relation $S_{\min}(\mathcal{N}_1) = \min_{\rho_A} S(\mathcal{N}_1(\rho_A))$, where $\rho_A$ and $\mathcal{N}_1(\rho_A)$ are the input and the output of the channel, while $S_{\min}(\mathcal{N}_1)$ is taken for the channel itself (leaving the input and output) [20]. It also follows from this relationship that to find the minimal output entropy of channel $\mathcal{N}_1$, the minimization can be made over the input system $\rho_A$. Let us assume that Alice puts the pure $|\psi\rangle$ state to the input of the completely depolarizing channel $\mathcal{N}_1$. In that case we have $S_{\min}(\mathcal{N}_1) = \min_{\rho_A} S(\mathcal{N}_1(\rho_A))$, where $\rho_A = |\psi\rangle\langle\psi|$ is a pure system. For this input the completely depolarizing qudit channel outputs the following state:

$$\mathcal{N}_{Compl.Depol.}(|\psi\rangle\langle\psi|) = \frac{1}{d}(\psi + I - \psi) = \frac{1}{d}I, \tag{28}$$

i.e., the channel outputs a maximally mixed state. The eigenvalues of the maximally mixed output state are independent from the channel input state $|\psi\rangle$, and they are equal to $e_1 = \frac{1}{d}$ and $e_2 = \frac{1}{d}d - 1 = 1 - \frac{1}{d}$. Using $\mathcal{N}_1 = \mathcal{N}_{Compl.Depol.}$, the minimum output entropy of the first quantum channel is

$$\begin{aligned} S_{\min}(\mathcal{N}_1) &= -\frac{1}{d}\log_2\left(\frac{1}{d}\right) - (d-1)\frac{1}{d}\log_2\left(\frac{1}{d}\right) \\ &= -\frac{1}{d}\log_2\left(\frac{1}{d}\right) + \frac{1}{d}\log_2\left(\frac{1}{d}\right) = 0. \end{aligned} \tag{29}$$

To describe the Holevo information of $\mathcal{N}_1$, we assume that Alice feeds the channel input system

$$\rho_{XA} = \sum_x p_X(x)|x\rangle\langle x|_X \otimes |\psi_x\rangle\langle\psi_x|_A, \tag{30}$$

where $X$ is Alice's classical register, $A$ is the input of the channel, $|\psi_x\rangle$ is the quantum system according to classical variable $x$ of the register $X$ fed to the input $A$ of the channel $\mathcal{N}_1$. The channel output of the first channel is written by system state

$$\sigma_{XBE} = U_{\mathcal{N}_1}^{A \to BE}(\rho_{XA}), \tag{31}$$

where $U_{\mathcal{N}_1}^{A \to BE}$ is the isometric extension of the completely depolarizing channel $\mathcal{N}_1$, $B$ is the output and $E$ is the environment of the channel [16] [20].

Assuming the input system $\rho_{XA}$, the Holevo information $\chi(\mathcal{N}_1)$ of the first channel can be expressed as

$$\begin{aligned}
C(\mathcal{N}_1) = \chi(\mathcal{N}_1) &= \max_{\rho^{XA}} I(X:B) \\
&= \max_{\rho^{XA}} (H(B)) - H(B|X) \\
&= \max_{\rho^{XA}} (H(B)) - H(E|X) \\
&= \log_2 d - \sum_x p_X(x) H(B) \\
&\leq \log_2 d - \min_{\rho_A} S(\mathcal{N}_1(\rho_A)) = \log_2 d - S_{\min}(\mathcal{N}_1),
\end{aligned} \tag{32}$$

where $I(X:B)$ is the quantum mutual information function between Alice's register $X$ and the channel output $B$. For pure input system $\rho_{XA}$, $H(B|X) = H(E|X)$ also follows. Generally $S_{\min}(\mathcal{N}_1) \leq \min_{\rho_A} S(\mathcal{N}_1(\rho_A))$, which is trivially equal for the completely depolarizing channel, i.e. for the classical capacity of the channel,

$$\begin{aligned}
C(\mathcal{N}_1) &= \chi(\mathcal{N}_1) \\
&= \log_2 d - \min_{\rho_A} S(\mathcal{N}_1(\rho_A)) \\
&= \log_2 d - S_{\min}(\mathcal{N}_1) = 0.
\end{aligned} \tag{33}$$

Since the first channel generates a maximally mixed output state, it also follows that any von Neumann measurement in the basis of the input state chosen from ensemble $\left\{\frac{1}{d}, |x\rangle\langle x|\right\}$, resulting $|y\rangle$, the probability of measuring $|x\rangle$ for a given output $|y\rangle$ is $p(y|x) = \frac{1}{d}$. For the conditional entropy $H(Y|X)$ we get

$$\begin{aligned}
H(Y|X) &= -\frac{1}{d}\log_2\left(\frac{1}{d}\right) - (d-1)\frac{1}{d}\log_2\left(\frac{1}{d}\right) \\
&= -\frac{1}{d}\log_2\left(\frac{1}{d}\right) + \frac{1}{d}\log_2\left(\frac{1}{d}\right) = 0.
\end{aligned} \tag{34}$$

*End of Proof of Lemma 1.1.* ∎

*End of the First Part of Proof of Theorem 1.* ∎

*Second Part of Proof of Theorem 1.*

Now we give the second part of the proof, i.e., we prove that the classical capacity of the cloning quantum channel $\mathcal{N}_2$ in $\mathcal{M}$ is zero.

**Lemma 1.2.** The classical capacity of the second channel $\mathcal{N}_2$ (the cloning channel) of $\mathcal{M}$ for the input system $\sigma_B$ (which is the output of the first zero-capacity channel $\mathcal{N}_1$ of $\mathcal{M}$) is zero.

*Proof of Lemma 1.2.*

The input $B$ of the second channel $\mathcal{N}_2$ (the cloning channel) is a maximally mixed state, i.e., $\Omega = 0.5$ (see (A.4)), thus the input system of the cloning channel can be expressed as

$$\frac{1}{2}|0\rangle\langle 0|_X \otimes |\varphi\rangle\langle \varphi|_{PB}^{(0)} + \frac{1}{2}|1\rangle\langle 1|_X \otimes |\varphi\rangle\langle \varphi|_{PB}^{(1)}, \tag{35}$$

where $|\varphi\rangle\langle \varphi|_{PB}^{(0)}$ and $|\varphi\rangle\langle \varphi|_{PB}^{(1)}$ are the purification of the maximally mixed input state $\frac{1}{2}(|0\rangle\langle 0| + |1\rangle\langle 1|)$ and can be expressed as

$$|\varphi\rangle_{PB}^{(0)} = |\varphi\rangle_{PB}^{(1)} = |\Phi_{00}\rangle = \frac{1}{\sqrt{2}}|00\rangle + |11\rangle. \tag{36}$$

From the purification state (36), using

$$\begin{aligned}\rho_{PB}^{(0)} &= |\varphi\rangle\langle \varphi|_{PB}^{(0)} = |\Phi_{00}\rangle\langle \Phi_{00}|, \\ \rho_{PB}^{(1)} &= |\varphi\rangle\langle \varphi|_{PB}^{(1)} = |\Phi_{00}\rangle\langle \Phi_{00}|,\end{aligned} \tag{37}$$

the output systems $\rho_B^{(0)}$ and $\rho_B^{(1)}$ can be expressed with the partial trace operator $Tr_P(\cdot)$ as follows:

$$\begin{aligned}
\rho_B^{(0)} &= Tr_P\left(\rho_{PB}^{(0)}\right) = \Omega|0\rangle\langle 0|_B + \kappa|1\rangle\langle 1|_B = \frac{1}{2}|0\rangle\langle 0|_B + \frac{1}{2}|1\rangle\langle 1|_B, \\
\rho_B^{(1)} &= Tr_P\left(\rho_{PB}^{(1)}\right) = \kappa|0\rangle\langle 0|_B + \Omega|1\rangle\langle 1|_B = \frac{1}{2}|0\rangle\langle 0|_B + \frac{1}{2}|1\rangle\langle 1|_B,
\end{aligned} \qquad (38)$$

where $\Omega$ and $\kappa$ with relation $\Omega + \kappa = 1$ are the parameters corresponding to the state on which the entropies are evaluated. (For the definitions of these parameters see (53) and (54).) The $Tr_P\left(\rho_{PB}^{(i)}\right)$, $i = 0,1$ trace out the purification state $P$ from the density matrix. As follows from (38), for a maximally mixed input state $\Omega = \kappa = 0.5$, and the purification of this state is the maximally entangled Bell state as given in $\rho_{PB}^{(0)}$ and $\rho_{PB}^{(1)}$, see (37).

If $\Omega = 0$, then the classical capacity of the cloning channel $1 \to N$ can be expressed as (output is denoted by $O$):

$$C(\mathcal{N}_2) = \max I(X:O) = 1 - \log_2 N + \frac{1}{\Delta_N}\sum_{i=0}^{N} i \log_2 i, \qquad (39)$$

where $\Delta_N$ was expressed in (19).

On the other hand, (39) cannot be used if $\Omega > 0$. To see it, we give the isometric extension of the $1 \to N$ cloning channel $\mathcal{N}_2$:

$$\begin{aligned}
\rho_{BOD}^{(0)} &= \frac{1}{\sqrt{\Delta_N}}\left[\sum_{i=0}^{N-1}\left(\sqrt{\Omega}\sqrt{N-i}|0\rangle_B|i\rangle_O + \sqrt{\kappa}\sqrt{i+1}|1\rangle_B|i+1\rangle_O\right)|i\rangle_E\right], \\
\rho_{BOD}^{(1)} &= \frac{1}{\sqrt{\Delta_N}}\left[\sum_{i=0}^{N-1}\left(\sqrt{\kappa}\sqrt{N-i}|0\rangle_B|i\rangle_O + \sqrt{\Omega}\sqrt{i+1}|1\rangle_B|i+1\rangle_O\right)|i\rangle_E\right],
\end{aligned} \qquad (40)$$

where $E$ is the environment of the cloning channel. After the input state has been transmitted through the cloning channel $\mathcal{N}_2$, the density matrix of the output system becomes

$$\rho_{XBOD} = \frac{1}{2}\left(|0\rangle\langle 0|_X \otimes \rho_{BOD}^{(0)} + |1\rangle\langle 1|_X \otimes \rho_{BOD}^{(1)}\right). \qquad (41)$$

where $X$ is Alice's classical register, $B$ is the input of the cloning channel, and $O$ and $D$ are the two outputs of the channel. The density matrix $\rho_{XO}$ can be expressed with the help of parameters $\Delta_N$ and

$$\lambda_i(\Omega) = (N - 2i)\Omega + i \quad \text{for } 0 \leq i \leq N. \qquad (42)$$

Using channel output density matrices

$$\rho_O^{(0)} = \sum_{i=0}^{N} \frac{\lambda_i(\Omega)}{\Delta_N} |i\rangle\langle i|_O,$$
$$\rho_O^{(1)} = \sum_{i=0}^{N} \frac{\lambda_i(\Omega)}{\Delta_N} |N-i\rangle\langle N-i|_O \qquad (43)$$

and

$$\rho_O = \frac{1}{N+1} \sum_{i=0}^{N} |i\rangle\langle i|_O, \qquad (44)$$

the output system state $XO$ can be expressed as

$$\rho_{XO} = \frac{1}{2}\left(|0\rangle\langle 0|_X \otimes \rho_O^{(0)} + |1\rangle\langle 1|_X \otimes \rho_O^{(1)}\right). \qquad (45)$$

From this, the quantum mutual information between Alice's register $X$ and the channel output $O$ is

$$C(\mathcal{N}_2) = \max I(X:O) = H(O) - H(O|X) = \log_2(N+1) - H\left(\frac{\lambda_i(\Omega)}{\Delta_N}\right), \qquad (46)$$

where the output and the conditional entropy, assuming $1 \to N$, can be calculated as

$$H(O) = \log_2(N+1), \qquad (47)$$

$$H(O|X) = H\left(\frac{\lambda_i(\Omega)}{\Delta_N}\right), \qquad (48)$$

where $H$ is the Shannon entropy function.

As was shown in [2], if $\Omega > 0$, then the formula in (39) cannot be used to derive the classical capacity of the $1 \to N$ cloning channel, so instead of it, we have to use (46). From (46), and using the fact that for the maximally mixed input state $\Omega = 0.5$ (see (38)) the classical capacity of the $1 \to 2$ (and trivially for any $1 \to N$) cloning channel is zero,

$$H(O|X) = H\left(\frac{\lambda_i(\Omega)}{\Delta_2}\right) = H\left(\frac{(N-2i) \cdot 0.5 + i}{\frac{N(N+1)}{2}}\right), \quad 0 \leq i \leq 2 \qquad (49)$$

i.e., $I(X:O) = 0$, since, for example, in the case of the $1 \to 2$ channel, because $\Omega = 0.5$, the classical capacity is

$$C(\mathcal{N}_2) = H(O) - H(O|X) = \log_2(3) - H\left(\frac{1}{N+1}\right) = \log_2(3) - H\left(\frac{1}{3}\right) = 0, \quad (50)$$

which means that the correlation between Alice's register $X$ and the output $O$ is zero, and $C(\mathcal{N}_2) = 0$ in (46) is proven.

*End of Proof of Lemma 1.2.* ∎

*Corollary 1.1:* From Lemmas 1.1 and 1.2 of the proof, the following conclusion can be derived. The classical capacity of the quantum channel $\mathcal{M} = \mathcal{N}_1 \circ \mathcal{N}_2$, where $\mathcal{N}_1$ can be any quantum channel that generates maximally mixed output state and $\mathcal{N}_2$ is the cloning channel, the classical capacity is zero, i.e., $C(\mathcal{M}) = 0$.

*End of the Second Part of Proof of Theorem 1.* ∎

**Theorem 2.** *For two quantum channels $\mathcal{M}_1$ and $\mathcal{M}_2$ according to the channel model discussed previously and having classical capacities $C(\mathcal{M}_1) = C(\mathcal{M}_2) = 0$, the joint classical capacity can be positive, i.e., $C(\mathcal{M}_1 \otimes \mathcal{M}_2) > 0$.*

In the first part of the proof we show that the auxiliary input system has to be entangled for the quasi-superactivation of classical capacity. Then, in the second part we prove that it cannot be a maximally entangled input system.

*First Part of Proof of Theorem 2.*

$C(\mathcal{M}) = 0$, i.e., $C(\mathcal{M}_1) = C(\mathcal{M}_2) = 0$, is already proven in Theorem 1. Now we show that the classical capacity of the joint channel structure $\mathcal{M}_1 \otimes \mathcal{M}_2$ can be quasi-superactivated, which will result in positive joint classical capacity $C(\mathcal{M}_1 \otimes \mathcal{M}_2) > 0$.

In the first phase, Alice correlates her classical register $X$ with two qubits denoted by $A_1 A_2$, then she feeds these states to the inputs of channels $\mathcal{M}_1$ and $\mathcal{M}_2$. In each case, the channels $\mathcal{N}_1$

(which can be any arbitrary quantum channel that generates maximally mixed output state) produces outputs $B_1B_2$, whose states have no classical correlation with the register X. Channels $\mathcal{N}_1$ destroy every classical correlation between the inputs $A_1A_2$ and the classical register X.

In the next step Alice does the following. She feeds the outputs of the first zero-capacity channels $\mathcal{N}_1$ to the inputs $B_1B_2$ of the second channels $\mathcal{N}_2$ (the cloning channels), while for the second inputs $C_1C_2$ of $\mathcal{M}_1 \otimes \mathcal{M}_2$ she puts the following entangled state:

$$|\Psi\rangle = \alpha|00\rangle + \beta|11\rangle, \tag{51}$$

where $0 < \alpha^2 \leq 0.5$ and $\beta^2 = 1 - \alpha^2$, and $\alpha^2 + \beta^2 = 1$. (*Note:* $\alpha^2 \leq 0.5$ is a sufficient condition on $\alpha^2$, since as $\alpha^2 > 0.5$ the system will become overparametrized, however as physical resources the two cases $0 < \alpha^2 \leq 0.5$ and $0.5 \leq \alpha^2 < 1$ are completely identical.)

We introduce a new representation of (51) using parameters $\sqrt{\Omega}$ and $\sqrt{\kappa}$ as

$$|\Psi\rangle = \sqrt{\Omega}|00\rangle + \sqrt{\kappa}|11\rangle. \tag{52}$$

where $\Omega$ the *amount of noise* (the *noise-parameter* is analogue to the von Neumann entropy of the pure two-qubit input system $|\Psi\rangle$ - it is 0 if $|\Psi\rangle$ is a *product state*, while it takes its maximum (0.5) if $|\Psi\rangle$ is a maximally entangled system. *Note:* $|\Psi\rangle$ is a *pure* two-qubit system, i.e., this noise is not the "mixedness" of the system.) is described by

$$\Omega = \alpha^2. \tag{53}$$

From $\Omega$ the parameter $\kappa$ can be expressed as

$$\kappa = 1 - \Omega, \tag{54}$$

with relation $\Omega + \kappa = 1$, are the parameters corresponding to the system of $|\Psi\rangle$, on which the entropies are evaluated. The input system $|\Psi\rangle$ of $C_1C_2$ in density matrix interpretation can be expressed as

$$\rho_{C_1C_2} = |\Psi\rangle\langle\Psi| = \alpha^2|00\rangle\langle00| + \alpha\beta|00\rangle\langle11| + \beta\alpha|11\rangle\langle00| + \beta^2|11\rangle\langle11|, \tag{55}$$

where $0 < \alpha^2 \leq 0.5$, with relation $\alpha^2 + \beta^2 = 1$. Using these parameters, the system state $\rho_{C_1C_2}$ in (55) can be rewritten as

$$\rho_{C_1C_2} = |\Psi\rangle\langle\Psi| = \Omega|00\rangle\langle00| + \sqrt{\Omega}\sqrt{\kappa}|00\rangle\langle11| + \sqrt{\kappa}\sqrt{\Omega}|11\rangle\langle00| + \kappa|11\rangle\langle11|, \quad (56)$$

where $\Omega$ and $\kappa$ are the parameters corresponding to the state of $\rho_{C_1C_2}$ of input system $C_1C_2$. As we will show, if and only if Alice feeds a non-maximally entangled state $|\Psi\rangle$ to the second inputs $C_1C_2$ of the channels, then some classical correlation between outputs $O_1O_2$ of zero-capacity channels $\mathcal{M}_1$ and $\mathcal{M}_2$ and the classical register $X$ can be restored. As we will prove in Lemma 2.1, Lemma 2.2, Lemma 2.3 and Theorem 3, if Alice would feed a maximally entangled state to the inputs $C_1C_2$ of $\mathcal{M}_1 \otimes \mathcal{M}_2$, the quasi-superactivation would not work, and the quasi-superactivated classical capacity will be equal to zero.

**Lemma 2.1.** Using any zero-capacity quantum channel $\mathcal{N}_1$ with maximally mixed channel output and an $1 \to N$ cloning channel $\mathcal{N}_2$, the quasi-superactivation of classical capacity $C(\mathcal{M}_1 \otimes \mathcal{M}_2)$ of the joint structure $\mathcal{M}_1 \otimes \mathcal{M}_2$ can be achieved if and only if *noise* is added in the form of an entangled state, $\rho_{C_1C_2} = |\Psi\rangle\langle\Psi|$. The noise is represented by the entangled input system $\rho_{C_1C_2} = |\Psi\rangle\langle\Psi|$, which has to be a non-maximally entangled system, otherwise $C(\mathcal{M}_1 \otimes \mathcal{M}_2) = 0$.

*Proof of Lemma 2.1.*

First we prove that the remote outputs $O_1D_2$ and $O_2D_1$ of the joint construction $\mathcal{M}_1 \otimes \mathcal{M}_2$ will be entangled if and only if the local outputs $O_1D_1$ and $O_2D_2$ are simultaneously un-entangled. We also show that for every $1 \to N$ cloning channel, entanglement on inputs $C_1C_2$ is a required condition in the characterization of the cloning channel $\mathcal{N}_2$ for the quasi-superactivation of the classical capacity of the joint structure $\mathcal{M}_1 \otimes \mathcal{M}_2$.

Assuming $1 \to 2$ cloning channel, and an untangled arbitrary *pure* input state $|\psi\rangle = \alpha|0\rangle + \beta|1\rangle$ the local outputs $O_1D_1$ and $O_2D_2$ of $\mathcal{M}_1 \otimes \mathcal{M}_2$ will be entangled [4], i.e., the required condition on the local outputs will not be satisfied (the local outputs have to be un-

entangled, as we will see in (66) and (67)). We note, for any $N > 2$, the local outputs of $\mathcal{M}_1 \otimes \mathcal{M}_2$ will be un-entangled without entanglement in auxiliary inputs $C_1 C_2$, i.e., for product state inputs there also will be no correlation between the remote outputs $O_1 D_2$ and $O_2 D_1$, either [4] [18-19].

For a *mixed* product state, the local outputs $O_1 D_1$ and $O_2 D_2$ will be un-correlated for any $1 \to N$, however in this case the remote outputs $O_1 D_2$ and $O_2 D_1$ will not be correlated [4].

From these statements it follows that the quasi-superactivation of classical capacity of $\mathcal{M}_1 \otimes \mathcal{M}_2$ requires an *entangled* auxiliary input system $\rho_{C_1 C_2} = |\Psi\rangle\langle\Psi|$ in the case of any $1 \to N$ cloning channel $\mathcal{N}_2$ in the joint channel $\mathcal{M}_1 \otimes \mathcal{M}_2$, since with product input states the entanglement between remote outputs $O_1 D_2$ and $O_2 D_1$ and the separability of local outputs $O_1 D_1$ and $O_2 D_2$ cannot be achieved in the same time.

*End of Proof of Lemma 2.1.* ■

Next, we give the proof of these two previous statements on the local and remote outputs of joint channel construction $\mathcal{M}_1 \otimes \mathcal{M}_2$.

The proof uses the results of Peres-Horodecki theorem [18-19]. The output systems of $\mathcal{M}_1 \otimes \mathcal{M}_2$ can be described by the density matrices $\rho_{O_1 D_1}$ and $\rho_{O_2 D_2}$, and $\rho_{O_1 D_1 O_2 D_2}$. As we will show, if Alice chooses appropriate values of $\Omega$, $\kappa$ in (56), then it is possible to reach the following property of the outputs of the channels $\mathcal{M}_1$ and $\mathcal{M}_2$.

**Lemma 2.2.** Using $1 \to 2$ (or any $1 \to N$) cloning channel and very special values of $\Omega$ and $\kappa$ in the input system $\rho_{C_1 C_2}$, the remote channel outputs $O_1 D_2$ and $O_2 D_1$ of $\mathcal{M}_1 \otimes \mathcal{M}_2$ will be entangled, while $O_1 D_1$ and $O_2 D_2$ will be un-entangled at the same time. There is no any correlation among input systems $B_1 C_1$ or $B_2 C_2$, however a special amount of entanglement in the input of $C_1C_2$ makes it possible to quasi-superactivate the classical capacity of the joint structure $\mathcal{M}_1 \otimes \mathcal{M}_2$.

*Proof of Lemma 2.2.*

To prove this statement, we use the density matrix representation $\rho_{O_1 D_1} = \rho_{O_2 D_2}$ of the output of system $\mathcal{M}_1 \otimes \mathcal{M}_2$, assuming the $1 \to N$ cloning channel $\mathcal{N}_2$. The remote channel outputs $O_1 D_2$ $(O_2 D_1)$ of $\mathcal{M}_1 \otimes \mathcal{M}_2$ will be entangled if and only if the amount of the entanglement in the input system $\rho_{C_1 C_2}$ was fed to inputs $C_1 C_2$ of $\mathcal{M}_1 \otimes \mathcal{M}_2$ is chosen properly. (It can be derived in the same way for outputs $O_2 D_1$.) Since the cloning quantum channel produces symmetric outputs, the following relation holds between the output density matrices of $\mathcal{M}_1 \otimes \mathcal{M}_2$

$$\rho_{O_1 D_2} = \rho_{O_2 D_1} \tag{57}$$

and

$$\rho_{O_1 D_1} = \rho_{O_2 D_2}. \tag{58}$$

Let us assume the input system $\rho_{C_1 C_2} = \rho_{C_1} \otimes \rho_{C_2}$ is a *pure un-entangled* system. (The inputs $C_1 C_2$ of $\mathcal{M}_1 \otimes \mathcal{M}_2$ consist of the *pure* un-entangled single qubit states $\rho_{C_1}$ and $\rho_{C_2}$.) Assume we have the unknown single qubit $|\psi\rangle = \alpha|0\rangle + \beta|1\rangle$ on the inputs $C_1 C_2$ denoted by $|\psi\rangle_{C_1} \otimes |\psi\rangle_{C_2}$. In this case, the density matrices $\rho_{O_1 D_1}$ and $\rho_{O_2 D_2}$ of the local outputs $O_1 D_1$ and $O_2 D_2$ of $\mathcal{M}_1 \otimes \mathcal{M}_2$ using basis $\{|00\rangle, |01\rangle, |10\rangle, |11\rangle\}_{O_1 D_2, O_2 D_1}$ and for the number of cloned outputs we introduce $M = N - 1$, can be expressed as [4]

$$\rho_{O_1 D_1} = \rho_{O_2 D_2}$$

$$= \frac{1}{6} \begin{pmatrix} \frac{(3M+5)|\beta|^2 + (M-1)|\alpha|^2}{M+1} & \frac{\alpha^*\beta(M+3)}{M+1} & \frac{\alpha^*\beta(M+3)}{M+1} & 0 \\ \frac{\alpha\beta^*(M+3)}{M+1} & 1 & 1 & \frac{\alpha^*\beta(M+3)}{M+1} \\ \frac{\alpha\beta^*(M+3)}{M+1} & 1 & 1 & \frac{\alpha^*\beta(M+3)}{M+1} \\ 0 & \frac{\alpha\beta^*(M+3)}{M+1} & \frac{\alpha\beta^*(M+3)}{M+1} & \frac{(3M+5)|\alpha|^2 + (M-1)|\beta|^2}{M+1} \end{pmatrix}.$$

According to the Peres-Horodecki theorem [18-19], from the eigenvalues of the density matrix $\rho_{O_1 D_1} = \rho_{O_2 D_2}$, Alice can determine for which values $N$ the local outputs $O_1 D_1$ and $O_2 D_2$ of $\mathcal{M}_1 \otimes \mathcal{M}_2$ will be entangled or un-entangled. Alice can do this, since if she finds that at least one of the eigenvalues $e = \{e_1, e_2, e_3, e_4\}$ of $\rho_{O_1 D_1} = \rho_{O_2 D_2}$ is negative, she will know surely that the local outputs of $\mathcal{M}_1 \otimes \mathcal{M}_2$ ($O_1$ with $D_1$, and $O_2$ with $D_2$) are entangled [4].

After some calculations, the eigenvalues $e = \{e_1, e_2, e_3, e_4\}$ of the matrix $\rho_{O_1 D_1} = \rho_{O_2 D_2}$ are

$$e = \left\{ \frac{1}{6}, \frac{1}{6}, \frac{1}{3} + \frac{\sqrt{2(5 + 4(N-1) + (N-1)^2)}}{6(N)}, \frac{1}{3} - \frac{\sqrt{2(5 + 4(N-1) + (N-1)^2)}}{6(N)} \right\}. \quad (59)$$

From (59) it can be concluded that one of the eigenvalues from $e$ will be negative *if and only if* $N = 2$, i.e. the number of cloned outputs is $M = N - 1 = 1$. This result means that if we have an *pure* un-entangled input system $\rho_{C_1} \otimes \rho_{C_2}$ on $C_1 C_2$, and we have an $1 \to 2$ quantum cloning channel in $\mathcal{M}_1 \otimes \mathcal{M}_2$ then the local outputs $O_1 D_1$ and $O_2 D_2$ will be entangled, which makes it impossible to create entanglement between the remote outputs of $\mathcal{M}_1 \otimes \mathcal{M}_2$, $O_1 D_2$ and $O_2 D_1$ [4] [18-19].

As we have concluded from (59), the quasi-superactivation of the classical capacity of the joint structure $\mathcal{M}_1 \otimes \mathcal{M}_2$ with a *pure* un-entangled input system $\rho_{C_1 C_2}$ and a $1 \to 2$ quantum cloning channel, cannot be achieved (see (66) and (67)). For *N=2*, an entangled (*Note*: in case of an EPR input state each channel gets a *mixed* input state, since one half of an EPR-state taken individually is a *mixed* quantum system, however the EPR state in itself is a pure two-qubit system.) input system $\rho_{C_1C_2}$ is a required condition for the quasi-superactivation of $\mathcal{M}_1 \otimes \mathcal{M}_2$.

*End of proof of Lemma 2.2.* ∎

*End of the First Part of Proof of Theorem 2.* ∎

On the other hand, as we will show, in all other cases, (i.e., $1 \to N$, $N > 2$), entanglement in $\rho_{C_1C_2}$ is also a required condition for the quasi-superactivation of classical capacity, since without entangled inputs the remote outputs $O_1D_2$ and $O_2D_1$ of $\mathcal{M}_1 \otimes \mathcal{M}_2$ would not be correlated [4] [18-19]. Moreover, the entangled input system $\rho_{C_1C_2}$ has also to satisfy a new requirement: it has to be a *non-maximally* entangled system. We will prove it in the second part of the proof.

*Second Part of Proof of Theorem 2.*

We show that if Alice uses the $1 \to N$ cloning channel for $\mathcal{N}_2$ and she feeds an entangled system to the second inputs of the cloning channels in the joint structure $\mathcal{M}_1 \otimes \mathcal{M}_2$, then to achieve the quasi-superactivation of the classical capacity, the entangled system has to be a *non-maximally entangled* system.

Let us assume that we have an $1 \to 2$ cloning channel. The entangled input is fed to the two inputs of the cloning channels $\mathcal{N}_2$ of the structure $\mathcal{M}_1 \otimes \mathcal{M}_2$, depicted by $C_1C_2$. The density matrix of the entangled system $\rho_{C_1C_2}$ is given in (56). Using the Peres-Horodecki theorem, it can be shown that input system $\rho_{C_1C_2}$ in (56) *will be entangled* for all values of $0 < \Omega \leq \frac{1}{2}$, since in this case one of the two determinants $\{\mathbf{d}_1, \mathbf{d}_2\}$ of the input system $\rho_{C_1C_2}$ will be negative [18-19] (*Note*: Parameter $\Omega$ is equal to 0, if and only if $\alpha^2 = 0, \beta^2 = 1$ (or if $\alpha^2 = 1, \beta^2 = 0$) which trivially leads to an un-entangled input system.). To prove it, we give the determinants of the input system $\rho_{C_1C_2}$ [4] [18-19]

$$\mathbf{d}_1 = \det \begin{bmatrix} \rho^{T_B}_{00,00} & \rho^{T_B}_{00,01} & \rho^{T_B}_{00,10} \\ \rho^{T_B}_{01,00} & \rho^{T_B}_{01,01} & \rho^{T_B}_{01,10} \\ \rho^{T_B}_{10,00} & \rho^{T_B}_{10,01} & \rho^{T_B}_{10,10} \end{bmatrix}, \quad \mathbf{d}_2 = \det\left[\rho^{T_B}\right], \tag{60}$$

(*Note*: the fourth row and columns are trivially omitted from (60), since for these matrices the required condition on the positivity is not satisfied, since the principal minors of these omitted matrices would not be positive.) where density operator $\rho_{ab,cd}$ describes a two qubit system $A$ and $B$ on the Hilbert space $\mathcal{H}_A \otimes \mathcal{H}_B$, as

$$\rho_{ab,cd} = \langle r_a | \langle t_b | \rho_0 | r_c \rangle | t_d \rangle, \tag{61}$$

and $\{|r_0\rangle = |0\rangle, |r_1\rangle = |1\rangle\}$, $\{|t_0\rangle = |0\rangle, |t_1\rangle = |1\rangle\}$ are the orthonormal bases of the first and second qubits, and $T$ denotes *the partial transpose* with respect to quantum system $B$, which can be expressed as

$$\rho^{T_B}_{ab,cd} = \rho_{ad,cb}. \tag{62}$$

(*Note*: For example, assuming a general density matrix $\rho = \sum_{ijkl} p^{ij}_{kl} |i\rangle\langle j| \otimes |k\rangle\langle l|$, which describes two quantum systems $A$ and $B$ in the Hilbert space $\mathcal{H}_A \otimes \mathcal{H}_B$, then the partial transpose $\rho^{T_B}$ with respect to $B$ system is

$$\rho^{T_B} = I \otimes T(\rho) = \sum_{ijkl} p^{ij}_{kl} |i\rangle\langle j| \otimes \left(|k\rangle\langle l|\right)^T = \sum_{ijkl} p^{ij}_{kl} |i\rangle\langle j| \otimes \left(|l\rangle\langle k|\right), \tag{63}$$

where $I$ is the identity transformation [18-19]. In matrix representation the $T_B$ partial transpose with respect to $B$ on density matrix $\rho$ can be expressed in the form of block matrix [3-4] [18-19]

$$\rho = \begin{pmatrix} a_{11} & a_{12} & \cdots & a_{1n} \\ a_{21} & a_{22} & & \\ \vdots & & \ddots & \\ a_{n1} & & & a_{nn} \end{pmatrix} \xrightarrow{I \otimes T(\rho)} \rho^{T_B} = \begin{pmatrix} a^T_{11} & a^T_{12} & \cdots & a^T_{1n} \\ a^T_{21} & a^T_{22} & & \\ \vdots & & \ddots & \\ a^T_{n1} & & & a^T_{nn} \end{pmatrix}, \tag{64}$$

where each block $a_{ij}$ is a square matrix with dimension $\dim \mathcal{H}_B$, while $n = \dim \mathcal{H}_A$. If the partial transpose matrix $\rho^{T_B}$ has a negative eigenvalue, then the quantum system $\rho$ is entangled [18-19].

The same results can be obtained if the partial transpose is taken with respect to system $A$, since $\rho^{T_A} = \left(\rho^{T_B}\right)^T$.)

From (62), the sub-determinants are [18-19]

$$\mathbf{d}_3 = \rho^{T_B}_{00,00}, \quad \mathbf{d}_4 = \rho^{T_B}_{00,00}\rho^{T_B}_{01,01} - \rho^{T_B}_{00,01}\rho^{T_B}_{01,00}. \tag{65}$$

After some calculations from the first two *determinants* $\{\mathbf{d}_1, \mathbf{d}_2\}$ one proves to be negative (while the sub-determinants $\{\mathbf{d}_3, \mathbf{d}_4\}$ are positive), which means that the necessary and sufficient condition for the separability of system $\rho_{C_1 C_2}$ is not satisfied. According to the Peres-Horodecki theorem, this condition is also equivalent to the necessary and sufficient condition that states that at least one of the eigenvalues of the partially transposed operator $\rho^T_{C_1 C_2}$ has to be negative, otherwise the input system $\rho_{C_1 C_2}$ will not be entangled [4]. (Or in other words, if all the eigenvalues are positive then the system is un-entangled).

Using an entangled system (56), one of the determinants from (60) will be negative, which proves that the input system fed by Alice to the inputs $C_1 C_2$ is entangled. We would like to see which special input conditions of this entangled system could help in the quasi-superactivation of the classical capacity of the joint structure $\mathcal{M}_1 \otimes \mathcal{M}_2$.

We prove that Alice (assuming $N=2$) can tune the values of $\Omega$ and $\kappa$ in (56) in that way, which leads to the quasi-superactivation of the classical capacity of the joint structure $\mathcal{M}_1 \otimes \mathcal{M}_2$. After Alice has fed the entangled input system $\rho_{C_1 C_2}$ to the inputs $C_1 C_2$, the output density matrices describing local outputs $O_1 D_1$, $O_2 D_2$ and remote outputs $O_1 D_2$, $O_2 D_1$ of $\mathcal{M}_1 \otimes \mathcal{M}_2$ can be expressed as follows [4]:

$$\rho_{O_1 D_1} = \rho_{O_2 D_2} = \frac{2\alpha^2}{3}|00\rangle\langle 00| + \frac{1}{6}(|01\rangle\langle 01| + |10\rangle\langle 10|) + \frac{2\beta^2}{3}|11\rangle\langle 11|, \tag{66}$$

and

$$\begin{aligned}\rho_{O_1 D_2} = \rho_{O_2 D_1} = &\frac{24\alpha^2 + 1}{36}|00\rangle\langle 00| + \frac{24\beta^2 + 1}{36}|11\rangle\langle 11| \\ &+ \frac{5}{36}(|01\rangle\langle 01| + |10\rangle\langle 10|) + \frac{4\alpha\beta}{9}(|00\rangle\langle 11| + |11\rangle\langle 00|).\end{aligned} \quad (67)$$

Here we can see an interesting thing. The results in (66) and (67) mean that the properties of the output density matrices $\rho_{O_1 D_1} = \rho_{O_2 D_2}$ and $\rho_{O_1 D_2} = \rho_{O_2 D_1}$ depend on the characterization of input system (56), which contradicts the original definition of the quantum cloning channel [4], since for a single cloning channel $\mathcal{N}_2$, the outputs of the channel are independent from the input state. (On the other hand, if we put these channels into our joint channel structure $\mathcal{M}_1 \otimes \mathcal{M}_2$ and feed entanglement to the inputs $C_1 C_2$, this statement does not remain true anymore.).

Alice's goal is to find those $\Omega$ and $\kappa$ values (see (53) and (54)) for which the density matrices $\rho_{O_1 D_2}$ and $\rho_{O_2 D_1}$ will be entangled—or, in other words, for which at least one determinant of the channel output density matrices $\rho_{O_1 D_2}$ and $\rho_{O_2 D_1}$ will be negative. The negativity of the determinants will hold if and only if Alice characterizes the entangled input system $\rho_{C_1 C_2}$ (see (56)) with the following values of $\Omega$:

$$\frac{1}{2} - \frac{\sqrt{39}}{16} \leq \Omega \leq \frac{1}{2} \ . \quad (68)$$

According to the Peres-Horodecki theorem, in these cases at least one determinant of each of the density matrices $\rho_{O_1 D_2}$ and $\rho_{O_2 D_1}$ will be negative, which proves that the outputs $O_1 D_2$ and $O_2 D_1$ of the joint channel construction $\mathcal{M}_1 \otimes \mathcal{M}_2$ will be entangled [4]. As we will show in the next step, the quasi-superactivation of the joint structure $\mathcal{M}_1 \otimes \mathcal{M}_2$ requires the following condition, namely $\Omega \neq \frac{1}{2}$, i.e., $\Omega$ cannot be equal to 0.5.

Now we prove that for some values of $\Omega$ and $\kappa$, the *local* outputs $O_1 D_1$ and $O_2 D_2$ will be simultaneously un-entangled, and the *remote* channel outputs $O_1 D_2$ and $O_2 D_1$ of $\mathcal{M}_1 \otimes \mathcal{M}_2$ will be entangled. Using again the Peres-Horodecki theorem [18-19], from the negativity of determinants of (60), it follows that for input system $\rho_{C_1 C_2} = \Omega |00\rangle\langle 00| + \kappa |11\rangle\langle 11|$ the local outputs $O_1 D_1$ and $O_2 D_2$ will be un-entangled [4-5] [18-19] if and only if

$$\frac{1}{2} - \frac{\sqrt{48}}{16} \leq \Omega \leq \frac{1}{2}. \tag{69}$$

These results indicate that if the remote outputs $O_1 D_2$ and $O_2 D_1$ of $\mathcal{M}_1 \otimes \mathcal{M}_2$ are entangled, there is no entanglement between the two local outputs $O_1 D_1$ and $O_2 D_2$, and these two properties hold simultaneously.

The results in (68) and (69) show that the properties of the entangled input system $\rho_{C_1 C_2}$ have to be chosen properly. The fact that it cannot be a maximally entangled state, i.e., $\Omega \neq \frac{1}{2}$, was already proven in (50), since it would destroy every possible classical capacity.

*Corollary 2.1*:

If Alice chose properly the amount of entanglement in the auxiliary input system $\rho_{C_1 C_2}$, she can realize entanglement between the remote channel outputs $O_1 D_2$ and $O_2 D_1$, while the local outputs $O_1 D_1$ and $O_2 D_2$ of $\mathcal{M}_1 \otimes \mathcal{M}_2$ will be un-entangled at the same time. From (68) and (69) we have concluded that the remote outputs of the joint channel construction $\mathcal{M}_1 \otimes \mathcal{M}_2$ can be entangled if and only the local outputs $O_1 D_1$ and $O_2 D_2$ are un-entangled.

We have already mentioned in (60) that for the entangled input system $\rho_{C_1 C_2}$, one of the eigenvalues of the transpose system $\rho^T_{C_1 C_2}$ will be negative and the entanglement among the channel outputs depends on the input system. We have also seen that for $\frac{1}{2} - \frac{\sqrt{39}}{16} \leq \Omega \leq \frac{1}{2}$, the outputs $O_1 D_2$ and $O_2 D_1$ will be entangled. Now, we prove the entanglement between outputs $O_1 D_2$ and $O_2 D_1$ of the joint construction $\mathcal{M}_1 \otimes \mathcal{M}_2$ in a different way.

**Lemma 2.3.** Feeding a non-maximally entangled, i.e., $\Omega \neq \frac{1}{2}$, auxiliary input system $\rho_{C_1 C_2}$ to inputs $C_1 C_2$ of the joint structure $\mathcal{M}_1 \otimes \mathcal{M}_2$, one eigenvalue of the output density matrix $\rho_{O_1 D_2} = \rho_{O_2 D_1}$ of the joint channel structure $\mathcal{M}_1 \otimes \mathcal{M}_2$ will be negative and the joint classical capacity will be positive.

*Proof of Lemma 2.3.*

In this proof we use again the Peres-Horodecki theorem. Feeding the entangled system $\rho_{C_1 C_2}$ to the inputs $C_1 C_2$, the output density matrix can be expressed as [4]

$$\rho_{O_1 D_2} = \rho_{O_2 D_1} = \sum_{i,j} \mathbf{M}_{i,j} \left| b_i \right\rangle \left\langle b_j \right| \Big|_{O_1 D_2 = O_2 D_1}, \qquad (70)$$

where $\left| b_1 \right\rangle = \left| 00 \right\rangle, \left| b_2 \right\rangle = \left| 01 \right\rangle, \left| b_3 \right\rangle = \left| 10 \right\rangle, \left| b_4 \right\rangle = \left| 11 \right\rangle$ and $\mathbf{M}$ is a $4 \times 4$ matrix

$$\mathbf{M} = \begin{pmatrix}
\sum_{mn} \begin{smallmatrix} \delta_{1,1}^{mn} \delta_{1,1}^{mn} + \delta_{2,1}^{mn} \delta_{2,1}^{mn} \\ +\delta_{1,3}^{mn} \delta_{1,3}^{mn} + \delta_{2,3}^{mn} \delta_{2,3}^{mn} \end{smallmatrix} & \sum_{mn} \begin{smallmatrix} \delta_{2,2}^{mn} \delta_{2,1}^{mn} + \delta_{1,2}^{mn} \delta_{1,1}^{mn} \\ +\delta_{1,4}^{mn} \delta_{1,3}^{mn} + \delta_{2,4}^{mn} \delta_{2,3}^{mn} \end{smallmatrix} & \sum_{mn} \begin{smallmatrix} \delta_{3,1}^{mn} \delta_{1,1}^{mn} + \delta_{4,1}^{mn} \delta_{2,1}^{mn} \\ +\delta_{3,3}^{mn} \delta_{1,3}^{mn} + \delta_{4,3}^{mn} \delta_{2,3}^{mn} \end{smallmatrix} & \sum_{mn} \begin{smallmatrix} \delta_{3,2}^{mn} \delta_{1,1}^{mn} + \delta_{4,2}^{mn} \delta_{2,1}^{mn} \\ +\delta_{3,4}^{mn} \delta_{1,3}^{mn} + \delta_{4,4}^{mn} \delta_{2,1}^{mn} \end{smallmatrix} \\
\sum_{mn} \begin{smallmatrix} \delta_{2,2}^{mn} \delta_{2,1}^{mn} + \delta_{1,2}^{mn} \delta_{1,1}^{mn} \\ +\delta_{1,4}^{mn} \delta_{1,3}^{mn} + \delta_{2,4}^{mn} \delta_{2,3}^{mn} \end{smallmatrix} & \sum_{mn} \begin{smallmatrix} \delta_{1,2}^{mn} \delta_{1,2}^{mn} + \delta_{2,2}^{mn} \delta_{2,2}^{mn} \\ +\delta_{1,4}^{mn} \delta_{1,4}^{mn} + \delta_{2,4}^{mn} \delta_{2,4}^{mn} \end{smallmatrix} & \sum_{mn} \begin{smallmatrix} \delta_{1,2}^{mn} \delta_{3,1}^{mn} + \delta_{2,2}^{mn} \delta_{4,1}^{mn} \\ +\delta_{1,4}^{mn} \delta_{3,3}^{mn} + \delta_{2,4}^{mn} \delta_{4,3}^{mn} \end{smallmatrix} & \sum_{mn} \begin{smallmatrix} \delta_{1,2}^{mn} \delta_{3,1}^{mn} + \delta_{2,2}^{mn} \delta_{4,2}^{mn} \\ +\delta_{1,4}^{mn} \delta_{3,4}^{mn} + \delta_{2,4}^{mn} \delta_{4,4}^{mn} \end{smallmatrix} \\
\sum_{mn} \begin{smallmatrix} \delta_{3,1}^{mn} \delta_{1,1}^{mn} + \delta_{4,1}^{mn} \delta_{2,1}^{mn} \\ +\delta_{3,3}^{mn} \delta_{1,3}^{mn} + \delta_{4,3}^{mn} \delta_{2,3}^{mn} \end{smallmatrix} & \sum_{mn} \begin{smallmatrix} \delta_{1,2}^{mn} \delta_{3,1}^{mn} + \delta_{2,2}^{mn} \delta_{4,1}^{mn} \\ +\delta_{1,4}^{mn} \delta_{3,3}^{mn} + \delta_{2,4}^{mn} \delta_{4,3}^{mn} \end{smallmatrix} & \sum_{mn} \begin{smallmatrix} \delta_{3,1}^{mn} \delta_{3,1}^{mn} + \delta_{4,1}^{mn} \delta_{4,1}^{mn} \\ +\delta_{3,3}^{mn} \delta_{3,3}^{mn} + \delta_{4,3}^{mn} \delta_{4,3}^{mn} \end{smallmatrix} & \sum_{mn} \begin{smallmatrix} \delta_{3,1}^{mn} \delta_{3,2}^{mn} + \delta_{4,1}^{mn} \delta_{4,2}^{mn} \\ +\delta_{3,3}^{mn} \delta_{3,4}^{mn} + \delta_{4,3}^{mn} \delta_{4,4}^{mn} \end{smallmatrix} \\
\sum_{mn} \begin{smallmatrix} \delta_{3,2}^{mn} \delta_{1,1}^{mn} + \delta_{4,2}^{mn} \delta_{2,1}^{mn} \\ +\delta_{3,4}^{mn} \delta_{1,3}^{mn} + \delta_{4,4}^{mn} \delta_{2,1}^{mn} \end{smallmatrix} & \sum_{mn} \begin{smallmatrix} \delta_{1,2}^{mn} \delta_{3,1}^{mn} + \delta_{2,2}^{mn} \delta_{4,2}^{mn} \\ +\delta_{1,4}^{mn} \delta_{3,4}^{mn} + \delta_{2,4}^{mn} \delta_{4,4}^{mn} \end{smallmatrix} & \sum_{mn} \begin{smallmatrix} \delta_{3,1}^{mn} \delta_{3,2}^{mn} + \delta_{4,1}^{mn} \delta_{4,2}^{mn} \\ +\delta_{3,3}^{mn} \delta_{3,4}^{mn} + \delta_{4,3}^{mn} \delta_{4,4}^{mn} \end{smallmatrix} & \sum_{mn} \begin{smallmatrix} \delta_{4,2}^{mn} \delta_{4,2}^{mn} + \delta_{3,2}^{mn} \delta_{3,2}^{mn} \\ +\delta_{3,4}^{mn} \delta_{3,4}^{mn} + \delta_{4,4}^{mn} \delta_{4,4}^{mn} \end{smallmatrix}
\end{pmatrix},$$

(71)

where $\delta$ is a mixture of the probability amplitudes $\alpha$ and $\beta$ of the entangled input in (51) weighted by the probability amplitudes $\varphi$ and $\gamma$ of the cloning quantum channel as [4]

$$\delta_{i,j}^{mn} = \alpha \varphi_m^i \varphi_n^j + \beta \gamma_m^i \gamma_n^j. \qquad (72)$$

The probability amplitudes $\varphi_m^i$ and $\gamma_m^i$ of the quantum cloning channel are assumed to be real and $\sum_{m=1}^{4} \left| \varphi_m^i \right|^2 = 1$, $\sum_{m=1}^{4} \left| \gamma_m^i \right|^2 = 1$, and $\sum_{m=1}^{4} \varphi_m^i \gamma_m^i = 1$. These probability amplitudes play an important role in the description of the state of the cloning quantum channel after the channel has transmitted its input, and the following relation holds:

$$|o_i\rangle = \sum_{m=1}^{4} \varphi_m^i |r_m\rangle,$$
$$|d_i\rangle = \sum_{m=1}^{4} \gamma_m^i |r_m\rangle,$$
(73)

where $|o_i\rangle$ and $|d_i\rangle$ are the two outputs of the quantum cloning channel in the basis of the four orthonormal basis states of the cloning quantum channel, denoted by $\{|r_m\rangle\}_{m=1}^{4}$.

According to the Peres-Horodecki theorem [18-19], the output density matrix $\rho_{C_1 D_2} = \rho_{C_2 D_1}$ of the joint channel construction $\mathcal{M}_1 \otimes \mathcal{M}_2$ will be entangled if and only if there is at least one eigenvalue of (70) is negative. From (71) and (72) it also follows that it would be possible if and only if input parameters $\Omega$ and $\kappa$ in (56) are between $\frac{1}{2} - \frac{\sqrt{39}}{16} \leq \Omega \leq \frac{1}{2}$. Otherwise, there are no exits probability amplitudes $\varphi_m^i$ and $\gamma_m^i$ in (72) for which at least one eigenvalue of (70) will be negative.

*End of Proof of Lemma 2.3.* ∎

*End of the Second Part of Proof of Theorem 2.* ∎

*Brief Summary*

In the proofs of Theorem 2 we have showed that the quasi-superactivation of the classical capacity requires entangled auxiliary input system. We have already proven that the input system cannot be a maximally entangled system. In the next theorem we prove that the amount of entanglement in the auxiliary input has to be chosen appropriately from a tight domain, otherwise the quasi-superactivation cannot be realized.

**Theorem 3.** *The classical capacity $C(\mathcal{M}_1 \otimes \mathcal{M}_2)$ of the joint channel structure $\mathcal{M}_1 \otimes \mathcal{M}_2$ will be positive if and only if the amount of entanglement in the auxiliary system is between $\frac{1}{2} - \frac{\sqrt{39}}{16} \leq \Omega < \frac{1}{2}$.*

*Proof.*

Let us assume that the required conditions do not hold, i.e., $0 \leq \Omega < \frac{1}{2} - \frac{\sqrt{39}}{16}$. In these cases, the density matrices $\rho_{O_1 D_2}$ and $\rho_{O_2 D_1}$ will be un-entangled, i.e., the outputs $O_1 D_2$ and $O_2 D_1$ of the joint channel $\mathcal{M}_1 \otimes \mathcal{M}_2$ will not be correlated. The outputs $O_1$ and $O_2$ of the joint channel $\mathcal{M}_1 \otimes \mathcal{M}_2$ are maximally mixed states, according to inputs $B_1$ and $B_2$. If there is no correlation between the two channels $\mathcal{M}_1$ and $\mathcal{M}_2$, these outputs are also maximally mixed states, which leads to

$$C(\mathcal{M}_1) = C(\mathcal{M}_2) = C(\mathcal{M}_1 \otimes \mathcal{M}_2) = 0. \tag{74}$$

The purifications of the maximally mixed inputs fed to $B_1$ and $B_2$ are the maximally entangled Bell states, as we have already seen in (36). It follows that if the entangled input system $\rho_{C_1 C_2}$ is characterized with different values of $\Omega$ and $\kappa$, as we stated in the theorem, there will be no entanglement between the remote outputs $O_1 D_2$ and $O_2 D_1$ of $\mathcal{M}_1 \otimes \mathcal{M}_2$, which trivially leads to zero classical capacity, since on both outputs $O_1 O_2$ a maximally mixed state will occur.

*Special case*

In case of $\Omega = \frac{1}{2}$ (i.e., assuming a maximally entangled input system in $C_1 C_2$) we have a special case. In this case, using the Peres-Horodecki theorem, we will find that remote outputs $O_1 D_2$ and $O_2 D_1$ of $\mathcal{M}_1 \otimes \mathcal{M}_2$ will be correlated and the local outputs remote outputs $O_1 D_1$ and $O_2 D_2$ of $\mathcal{M}_1 \otimes \mathcal{M}_2$ will be un-entangled; however, the quasi-superactivation of the classical capacity will not work, either. In this special case Alice's register $X$ and the channel outputs $O_1$ and $O_2$ will be completely uncorrelated, meaning that the classical capacity of the joint structure will be equal to zero, $C(\mathcal{M}_1 \otimes \mathcal{M}_2) = 0$.

The requirements for the amount of entanglement in the input system $\rho_{C_1C_2}$ for the quasi-superactivation are also summarized in Fig. 5.

**Fig. 5.** (a): If the amount of entanglement in the auxiliary input system of the joint channel is chosen inappropriately, then the channel construction cannot be used to transmit any classical information, and the quasi-superactivation effect will not occur. In this case, any classical correlation between register $X$ and channel outputs $O_1O_2$ will completely vanish.

(b): The classical capacity $C(\mathcal{M}_1 \otimes \mathcal{M}_2)$ of the joint channel structure $\mathcal{M}_1 \otimes \mathcal{M}_2$ will be positive if and only if the amount of entanglement in the auxiliary input system $C_1C_2$ is chosen from a very tight domain by Alice. In this case, classical correlation between register $X$ and outputs $O_1$ and $O_2$ will occur on the channel output.

*End of Proof of Theorem 3.* ∎

After these results we derive the amount of quasi-superactivated classical capacity of the joint structure $\mathcal{M}_1 \otimes \mathcal{M}_2$.

# 5 Results

After in Theorem 2 and Theorem 3 we have derived the necessary and sufficient conditions for the quasi-superactivation of the classical capacity of the joint structure, we give an explicit formula for the maximally achievable quasi-superactivated joint classical channel capacity $C(\mathcal{M}_1 \otimes \mathcal{M}_2)$. To determine it, we will use the formulas of (39) and (46).

As we have derived, the level of the quasi-superactivation of the classical capacity $C(\mathcal{M}_1 \otimes \mathcal{M}_2)$ of the joint structure $\mathcal{M}_1 \otimes \mathcal{M}_2$ depends on the properties of the auxiliary entangled input that was fed to inputs $C_1 C_2$ of the joint structure $\mathcal{M}_1 \otimes \mathcal{M}_2$. If a maximally entangled state was given to the inputs $C_1 C_2$, the quasi-superactivated capacity completely vanished. We proved that the quasi-superactivation of the classical capacity of the joint structure $\mathcal{M}_1 \otimes \mathcal{M}_2$ works if and only if the quantum noise is added in the form of a non-maximally entangled state.

First, we characterize the inputs of the cloning channels $\mathcal{N}_2$ from the joint structure $\mathcal{M}_1 \otimes \mathcal{M}_2$. The system state in terms of Alice's register $X$, the purification state $P$, and the input $B$ of the cloning channel $\mathcal{N}_2$ is

$$\frac{1}{2}|0\rangle\langle 0|_X \otimes \rho_{PB}^{(0)} + \frac{1}{2}|1\rangle\langle 1|_X \otimes \rho_{PB}^{(1)}, \tag{75}$$

where $\rho_{PB}^{(0)}$ and $\rho_{PB}^{(1)}$ are the purification of the input states $\rho_B^{(0)}$ and $\rho_B^{(1)}$. Using (36), the system states of $\rho_B^{(0)}$ and $\rho_B^{(1)}$ from $\rho_{PB}^{(0)}$ and $\rho_{PB}^{(1)}$ can be expressed with the partial trace operator $Tr_P\left(\rho_{PB}^{(i)}\right)$, $i = 0,1$ depending on $\alpha^2$ in (55), as follows:

$$\rho_B^{(0)} = Tr_P\left(\rho_{PB}^{(0)}\right) = \Omega|0\rangle\langle 0|_B + \kappa|1\rangle\langle 1|_B = \left(\frac{1}{2} - \frac{\sqrt{39}}{16} \leq \Omega < \frac{1}{2}\right)|0\rangle\langle 0|_B + \left(\frac{1}{2} < \kappa \leq \frac{1}{2} + \frac{\sqrt{39}}{16}\right)|1\rangle\langle 1|_B,$$

$$\rho_B^{(1)} = Tr_P\left(\rho_{PB}^{(1)}\right) = \kappa|0\rangle\langle 0|_B + \Omega|1\rangle\langle 1|_B = \left(\frac{1}{2} < \kappa \leq \frac{1}{2} + \frac{\sqrt{39}}{16}\right)|0\rangle\langle 0|_B + \left(\frac{1}{2} - \frac{\sqrt{39}}{16} \leq \Omega < \frac{1}{2}\right)|1\rangle\langle 1|_B,$$

$$\tag{76}$$

the input system of which, assuming $N = 2$ (the maximally achievable capacity decreases as $N$ increases, i.e., the best results on the classical capacity can be obtained if $N = 2$) leads us to classical capacity

$$C(\mathcal{M}_1 \otimes \mathcal{M}_2) = \max I(X:O) = H(O) - H(O|X) = \log_2(N+1) - H\left(\frac{\lambda_i(\Omega)}{\Delta_2}\right), \tag{77}$$

where the output and the conditional entropy can be expressed as

$$H(O) = \log_2(2+1), \tag{78}$$

$$H(O|X) = H\left(\frac{\lambda_i(\Omega)}{\Delta_2}\right) = H\left(\frac{(2-2i)\cdot\Omega + i}{\frac{2(2+1)}{2}}\right), \text{ for } 0 \leq i \leq 2, \tag{79}$$

which for $\frac{1}{2} - \frac{\sqrt{39}}{16} \leq \Omega < \frac{1}{2}$ leads to quasi-superactivated classical capacity $C(\mathcal{M}_1 \otimes \mathcal{M}_2)$:

$$\log_2(N+1) - H\left(\frac{\lambda_i\left(\frac{1}{2}\right)}{\Delta_N}\right) = \log_2(N+1) - H\left(\frac{\frac{1}{2}N}{\frac{N(N+1)}{2}}\right) = \log_2(2+1) - H\left(\frac{1}{2+1}\right) = 0$$

$$< C(\mathcal{M}_1 \otimes \mathcal{M}_2) \leq$$

$$\log_2(N+1) - H\left(\frac{\lambda_i\left(\frac{1}{2} - \frac{\sqrt{39}}{16}\right)}{\Delta_N}\right) = \log_2 3 - H\left(\frac{(2-2i)\cdot\left(\frac{1}{2} - \frac{\sqrt{39}}{16}\right) + i}{\frac{2(2+1)}{2}}\right) \text{ for } 0 \leq i \leq 2 \tag{80}$$

$$= \log_2 3 - H\left(\frac{2\cdot\left(\frac{1}{2} - \frac{\sqrt{39}}{16}\right)}{3}\right) - H\left(\frac{1}{3}\right) - H\left(\frac{2 - 2\cdot\left(\frac{1}{2} - \frac{\sqrt{39}}{16}\right)}{3}\right) = 0.3354,$$

where $\lambda_i(\Omega)$ and $\Delta_N$ were defined in (42) and (19). From (80), it follows that for $N=2$, the quasi-superactivated classical capacity of the joint structure $\mathcal{M}_1 \otimes \mathcal{M}_2$ for $\alpha^2 \leq 0.5$ is upper-bounded by

$$C(\mathcal{M}_1 \otimes \mathcal{M}_2) \leq \log_2 3 - H\left(\frac{\lambda_i(\Omega)}{\Delta_2}\right), \tag{81}$$

depending on $\Omega$. From (81) it follows that the level of the quasi-superactivation of the classical capacity $C(\mathcal{M}_1 \otimes \mathcal{M}_2)$ of the joint structure $\mathcal{M}_1 \otimes \mathcal{M}_2$ depends on the properties of the second entangled input that was fed to inputs $C_1C_2$ of the joint structure $\mathcal{M}_1 \otimes \mathcal{M}_2$. If a maximally entangled state (i.e., $\Omega = \frac{1}{2}$) was given to the inputs $C_1C_2$, the quasi-superactivated capacity completely vanished. Here, we have also used (50), in which it was proven that the quasi-superactivation of the classical capacity of the joint structure $\mathcal{M}_1 \otimes \mathcal{M}_2$ works if and only if $\Omega \neq \frac{1}{2}$, i.e., it cannot be a maximally entangled state.

The achievable quasi-superactivated joint classical capacity $C(\mathcal{M}_1 \otimes \mathcal{M}_2)$ of the joint structure $\mathcal{M}_1 \otimes \mathcal{M}_2$ (for $N = 2$) in function of the noise-parameter $\Omega$ is shown in Fig. 6.

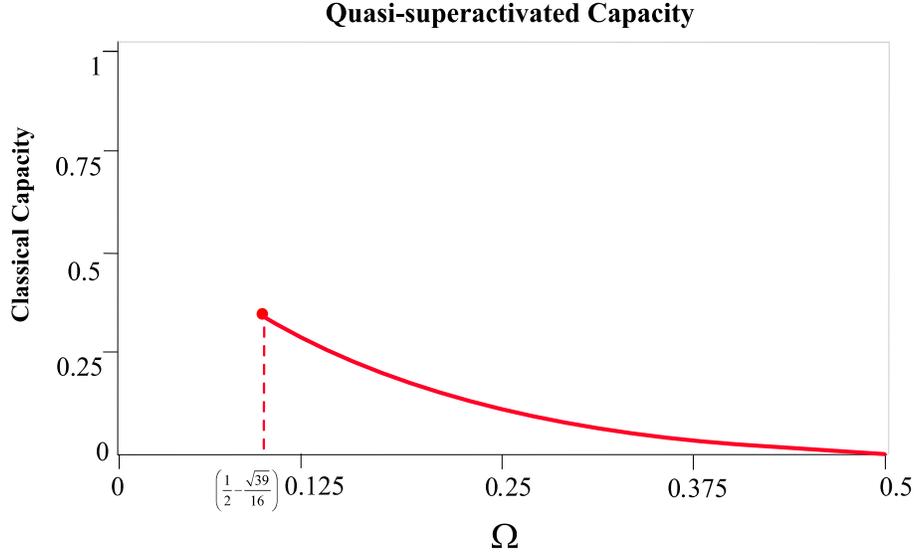

**Fig. 6.** The level of quasi-superactivated classical capacity of the joint structure $\mathcal{M}_1 \otimes \mathcal{M}_2$ depends on the amount of entanglement in the EPR input that was fed by Alice to the inputs of the joint channel structure. The quasi-superactivated quantum mutual information function takes its maximum if in the entangled auxiliary input system of $C_1 C_2$, $\frac{1}{2} - \frac{\sqrt{39}}{16} \leq \alpha^2 < 0.5$, while completely vanishes if $\alpha^2 = 0.5$.

As we have found, the level of quasi-superactivation depends on the amount of entanglement in the inputs $C_1 C_2$ of the joint structure $\mathcal{M}_1 \otimes \mathcal{M}_2$, and the classical capacity can be greater zero only in a well-specified strict domain.

# 6 Conclusions

Since the superactivation of the classical capacity of quantum channels is trivially not possible, before our work the transmission of classical information over zero-capacity quantum channels was also seemed to be not possible. In this paper the term quasi-superactivation is firstly introduced. We proved that by adding quantum entanglement to zero-capacity quantum channels, *classical* information transmission is possible. The quasi-superactivation is similar to the superactivation effect, thus positive capacity can be achieved with noisy quantum channels that were initially completely useless for classical communication. However, an important difference that quasi-superactivation is limited neither by any preliminary conditions of the originally introduced superactivation effect nor on the maps of other channels involved to the joint channel structure. As we have proven, besides that there exists zero-capacity quantum channels with positive quantum capacity, it is also possible to find zero-capacity quantum channels with individually zero classical capacities, which if employed in a joint channel construction can transmit classical information. We hope our results help to reveal the strange and mysterious world of quantum information, and to characterize and exploit the hidden possibilities in information transmission over quantum channels in the communication systems and networks of the future. The proposed scheme uses only very simple elements, which allows for a very effective implementation and verification in practice.

In future work we would like to analyze the possibilities in the qudit quantum cloners [28]. As we hope further superactivated capacity-regions can be opened by the exploitation of qudit cloners and the superactivated joint channel capacity can be increased even more [28]. It should also be an interesting question that for what dimension of qudits the superactivation reaches its maximum. Finally, based on the results of Bradler et al. [27], we would like to extend the proposed method to the superactivation of asymptotic quantum capacity of zero-capacity quantum channels.

## Acknowledgements

LGY would like to thank Kamil Bradler for useful discussions and comments. The results discussed above are supported by the grant TAMOP-4.2.1/B-09/1/KMR-2010-0002, 4.2.2.B-10/1--2010-0009 and COST Action MP1006.

# Supplementary Information

## A.1 Purification

The purification gives us a new viewpoint on the noise of the quantum channel. Assuming Alice's side $A$ and Alice's classical register $X$, the spectral decomposition of the density operator $\rho_A$ can be expressed as

$$\rho_A = \sum_x p_X(x) |x\rangle\langle x|_A, \tag{A.1}$$

where $p_X(x)$ is the probability of variable $x$ in Alice's classical register $X$. The $\{p_X(x), |x\rangle\}$ together is called an ensemble, where $|x\rangle$ is a quantum state according to classical variable $x$. Using the set of orthonormal basis vectors $\{|x\rangle_P\}_{x \in X}$ of the purification system $P$, the purification of (A.1) can be given in the following way:

$$|\varphi\rangle_{PA} = \sum_x \sqrt{p_X(x)} |x\rangle_P |x\rangle_A. \tag{A.2}$$

Using parameters $\Omega$ and $\kappa$, (A.2) can be rewritten as

$$|\varphi\rangle\langle\varphi|_{PB} = \Omega|00\rangle\langle 00| + \sqrt{\Omega}\sqrt{\kappa}|00\rangle\langle 11| + \sqrt{\kappa}\sqrt{\Omega}|11\rangle\langle 00| + \kappa|11\rangle\langle 11|. \tag{A.3}$$

From (7) we know that the purification of the maximally mixed state is the Bell state $|\Phi_{00}\rangle$, i.e., from the purified system state $BP$, the original system state $B$ can be expressed as

$$\sigma_B = Tr_P\left(|\varphi\rangle\langle\varphi|_{PB}\right) = \Omega|0\rangle\langle 0| + \kappa|1\rangle\langle 1| = \frac{1}{2}\left(|0\rangle\langle 0| + |1\rangle\langle 1|\right), \tag{A.4}$$

thus in (A.4) we have $\Omega = \kappa = \frac{1}{2}$. In (9), the noise of the zero-capacity channel $\mathcal{N}_1$ results in a maximally entangled state between the output system $\sigma_B$ and the environment (i.e., the purification system $P$). From the purified system state $|\varphi\rangle_{PA}$, the original system state $\rho_A$ can be expressed with the partial trace operator $Tr_P(\cdot)$, which operator traces out the purification state (i.e., the environment) from the system

$$\rho_A = Tr_P\left(|\varphi\rangle\langle\varphi|_{PA}\right). \tag{A.5}$$

From joint system (A.2) and the purified state (A.5), one can introduce a new definition. The *extension* of $\rho_A$ can be given as [20]

$$\rho_A = Tr_P\left(\omega_{PA}\right), \tag{A.6}$$

where $\omega_{PA}$ is a noisy quantum system.

## A.2 Partial Trace

If we have a density matrix which describes only a subset of a larger quantum space, then we talk about the reduced density matrix. The larger quantum system can be expressed as the tensor product of the reduced density matrices of the subsystems, if there is no correlation (entanglement) between the subsystems [16] [20]. On the other hand, if we have two subsystems with reduced density matrices $\rho_A$ and $\rho_B$, then from the overall density matrix denoted by $\rho_{AB}$ the subsystems can be expressed as

$$\rho_A = Tr_B\left(\rho_{AB}\right) \text{ and } \rho_B = Tr_A\left(\rho_{AB}\right), \tag{A.7}$$

where $Tr_B$ and $Tr_A$ refers to the partial trace operators. So, this partial trace operator can be used to generate one of the subsystems from the joint state $\rho_{AB} = |\psi_A\rangle\langle\psi_A| \otimes |\psi_B\rangle\langle\psi_B|$, then

$$\begin{aligned}\rho_A &= Tr_B\left(\rho_{AB}\right) = Tr_B\left(|\psi_A\rangle\langle\psi_A| \otimes |\psi_B\rangle\langle\psi_B|\right) \\ &= |\psi_A\rangle\langle\psi_A|Tr\left(|\psi_B\rangle\langle\psi_B|\right) = |\psi_A\rangle\langle\psi_A|\langle\psi_B|\psi_B\rangle.\end{aligned} \tag{A.8}$$

Since the inner product is trivially $\langle\psi_B|\psi_B\rangle = 1$, therefore

$$Tr_B(\rho_{AB}) = \langle\psi_B|\psi_B\rangle|\psi_A\rangle\langle\psi_A| = |\psi_A\rangle\langle\psi_A| = \rho_A. \tag{A.9}$$

In the calculation, we used the fact that $Tr(|\psi_1\rangle\langle\psi_2|) = \langle\psi_2|\psi_1\rangle$. In general, if we have to systems $A = |i\rangle\langle k|$ and $B = |j\rangle\langle l|$, then the partial trace can be calculated as

$$Tr_B(A \otimes B) = A\,Tr(B), \tag{A.10}$$

since

$$\begin{aligned}Tr_2(|i\rangle\langle k| \otimes |j\rangle\langle l|) &= |i\rangle\langle k| \otimes Tr(|j\rangle\langle l|) \\ &= |i\rangle\langle k| \otimes \langle l|j\rangle \\ &= \langle l|j\rangle|i\rangle\langle k|,\end{aligned} \tag{A.11}$$

where $|i\rangle\langle k| \otimes |j\rangle\langle l| = |i\rangle|j\rangle(|k\rangle|l\rangle)^T$. In this expression we have used the fact that $(AB^T) \otimes (CD^T) = (A \otimes C)(B^T \otimes D^T) = (A \otimes C)(B \otimes D)^T$.

## A.3 Isometric Extension

Isometric extension is of the utmost importance for us, because it helps us to understand what happens between the quantum channel and its environment whenever a quantum state is transmitted from Alice to Bob. The isometric extension is an important tool in Quantum Information Theory, and it possesses especially deep relevance to the description of the quasi-superactivation of the classical capacity. The isometric extension of the quantum channel $\mathcal{N}$ is simply the unitary representation of the channel

$$\mathcal{N}: U_{A \to BE}, \tag{A.12}$$

where $A$ is the input system, $B$ is the channel output of the unitary transformation, and $E$ is the environment of the quantum channel. The isometric extension of a quantum channel $\mathcal{N}$ enables us to describe the transmission of quantum information over a noisy quantum channel in a "one-sender and two-receiver" view: the first receiver is Bob, and the second receiver is the environment of the channel. In other words, the output of the noisy quantum channel $\mathcal{N}$ can be described only after the environment of the channel is traced out. By applying $U_{A \to BE}$ to the input quantum system $\rho$, we have

$$Tr_E \left( U_{A \to BE} \left( \rho_A \right) \right) = \mathcal{N} \left( \rho_A \right). \tag{A.13}$$

Thus, we have presented the "extension" part. Now, let us see the "isometric" part. The isometry of a unitary map $U$ holds two attributes. First, $U^\dagger U = I$, that is, it behaves just like an ordinary unitary transformation, and $UU^\dagger = P_{BE}$, where $P_{BE}$ is a projector applied to the joint output $BE$ of the quantum channel.

## A.4 Kraus Representation

The map of the quantum channel can also be expressed with a special representation called the *Kraus Representation*. For a given input system $\rho_A$ and the quantum channel $\mathcal{N}$, this representation can be expressed as

$$\mathcal{N} \left( \rho_A \right) = \sum_i N_i \rho_A N_i^\dagger, \tag{A.14}$$

where $N_i$ are the Kraus operators, and $\sum_i N_i^\dagger N_i = I$. The isometric extension of $\mathcal{N}$ by means of the *Kraus Representation* can be expressed as

$$\mathcal{N} \left( \rho_A \right) = \sum_i N_i \rho_A N_i^\dagger \to U_{A \to BE} \left( \rho_A \right) = \sum_i N_i \otimes \left| i \right\rangle_E. \tag{A.15}$$

The action of the quantum channel $\mathcal{N}$ on an operator $\left| k \right\rangle \left\langle l \right|$, where $\left\{ \left| k \right\rangle \right\}$ is an orthonormal basis, also can be given in operator form using the Kraus operator $N_{kl} = \mathcal{N} \left( \left| k \right\rangle \left\langle l \right| \right)$. By exploiting the property $UU^\dagger = P_{BE}$, for the input quantum system $\rho_A$

$$U_{A \to BE} \left( \rho_A \right) = U \rho_A U^\dagger = \left( \sum_i N_i \otimes \left| i \right\rangle_E \right) \rho_A \left( \sum_j N_j^\dagger \otimes \left\langle j \right|_E \right) = \sum_{i,j} N_i \rho_A N_j^\dagger \otimes \left| i \right\rangle \left\langle j \right|_E. \tag{A.16}$$

Now, let us apply (A.13). If we trace out the environment, we get

$$Tr_E\left(U_{A\to BE}\left(\rho_A\right)\right) = \sum_i N_i \rho_A N_i^\dagger, \qquad (A.17)$$

which represents exactly the same conditions as (A.15).

## A.5 Quantum Channel Concatenation

The output of the concatenated $\mathcal{M} = \mathcal{N}_1 \circ \mathcal{N}_2$ can be described formally in the following way. For the input density matrix $\rho$, the output of a quantum channel $\mathcal{N}_1$ and $\mathcal{N}_2$ can be given by its Kraus representation as $\mathcal{N}_1(\rho) = \sum_i N_i^{(1)} \rho N_i^{(1)\dagger}$ and $\mathcal{N}_2(\rho) = \sum_i N_i^{(2)} \rho N_i^{(2)\dagger}$. The output of the concatenated structure of two quantum channels is simply

$$\mathcal{N}_2 \circ \mathcal{N}_1(\rho) = \sum_i N_i^{(2)} \mathcal{N}_1(\rho) N_i^{(2)\dagger} = \sum_{i,i'} N_i^{(2)} N_{i'}^{(1)} \rho N_{i'}^{(1)\dagger} N_i^{(2)\dagger}, \qquad (A.18)$$

where $\mathrm{K} = \left\{N_i^{(1)}\right\}, \left\{N_i^{(2)}\right\}$ are the Kraus operators of the individual channels, while $\mathrm{K} = \left\{N_i^{(1)}, N_i^{(2)}\right\}_{i,i'}$ are the Kraus operators of the concatenated channels $\mathcal{N}_2 \mathcal{N}_1$.